\documentclass[aps,pra,reprint,showpacs,superscriptaddress,floatfix]{revtex4-1}
\usepackage{units}
\usepackage{amsmath}
\usepackage{amssymb}
\usepackage[dvips]{graphicx}
\usepackage{textcomp}
\usepackage{color}
\usepackage{hyperref}
\begin{document}
\title{Topological tight-binding models from non-trivial square roots}

\author{J. Arkinstall}
\affiliation{Department of Physics, Lancaster University, Lancaster, LA1 4YB, United Kingdom}

\author{M. H. Teimourpour}
\affiliation{Department of Physics and Henes Center for Quantum Phenomena, Michigan Technological University, Houghton, Michigan 49931, USA}

\author{L. Feng}
\affiliation{Department of Electrical Engineering, University at Buffalo, The State University of New York, Buffalo, NY 14260, USA}

\author{R. El-Ganainy}
\email{ganainy@mtu.edu}
\affiliation{Department of Physics and Henes Center for Quantum Phenomena, Michigan Technological University, Houghton, Michigan 49931, USA}

\author{H. Schomerus}
\email{h.schomerus@lancaster.ac.uk}
\affiliation{Department of Physics, Lancaster University, Lancaster, LA1 4YB, United Kingdom}

\date{\today}

\begin{abstract}
We describe a versatile mechanism that provides tight-binding models with an enriched, topologically nontrivial bandstructure.
The mechanism is algebraic in nature, and leads to tight-binding models that can be interpreted as a non-trivial square root of a parent lattice Hamiltonian---in analogy to the passage from a Klein-Gordon equation to a Dirac equation. In the tight-binding setting, the square-root operation admits to induce spectral symmetries at the expense of broken crystal symmetries.
As we illustrate in detail for a simple one-dimensional example, the emergent and inherited spectral symmetries
equip the energy gaps with independent topological quantum numbers
that control the formation of topologically protected states.
We also describe an implementation of this system in silicon photonic structures,  outline applications in higher dimensions, and provide a general argument for the origin and nature of the emergent symmetries, which are typically nonsymmorphic.
\end{abstract}

\pacs{42.55.Sa, 42.55.Ah, 42.60.Da}

\maketitle

\section{Introduction}

As the story goes, in 1927 Niels Bohr asked Paul Dirac ``What are you working on Mr. Dirac?'' to which Dirac replied ``I'm trying to take the square root of something.'' Once Dirac achieved his goal, to identify the desired operator that squares to the Klein-Gordon equation, he had not only laid down a description of relativistic electrons replete with spin and antimatter \cite{Dirac1928,Dirac1958}. As it emerged later, Dirac's very same operator also plays a central role for topological considerations in differential geometry, where the Atiyah-Singer index theorem addresses its zero modes \cite{Atiyah1963}. The zero modes in the topological materials considered today are a direct extension of this connection \cite{Hasan2010,Qi2011}.
Fundamental symmetries can guarantee that all positive-energy states are  paired with  negative-energy states, with the exception of a protected set of zero-energy states whose  number $|\nu|$ is obtained from a topological invariant.
These properties may follow from  a charge-conjugation symmetry, as encountered in superconductors
\cite{Altland1997,Beenakker2015}, or from a chiral symmetry, as encountered for the Dirac operator
\cite{Verbaarschot1993,Verbaarschot2000}; both operations anticommute with the Hamiltonian and therefore single out a spectral symmetry point.
In combination with possible invariance under time-reversal, these spectral symmetries determine a ten-fold system of universality classes \cite{Zirnbauer2011,Beenakker2015}, which can be further extended by including aspects of dimensionality \cite{Ryu2010,Teo2010} and the space group (i.e., crystal symmetries) \cite{Chiu2016}---for example, nonsymmorphic symmetries involving fractional lattice translations can replace fundamental symmetries normally associated with fermionic systems \cite{Lu2016}.
  Depending on the universality class, the topological invariant may take the values
$\nu\in\{0,1\}$, leading to the notion of a $\mathbb{Z}_2$ invariant, or be any integer, leading to the notion of a $\mathbb{Z}$ invariant.
These topological features
are not present in the Klein-Gordon equation, from which Dirac had started to take the square-root of, a task which was non-trivial as it required him to introduce extra components and matrices.

Here we describe how rich topological effects arise when one takes an analogous non-trivial square root on a tight-binding lattice.  Tight-binding lattices provide an ubiquitous description of electronic bands in crystalline solids, but also extend to atoms and photons in suitably engineered optical and photonic lattices. This includes topological systems in all universality classes, such as the paradigmatic Su-Schrieffer-Heeger model, originally proposed for polyacetylene \cite{Su1979}, and non-topological variants such as the Rice-Mele model for conjugated polymers \cite{Rice1982},
both of which have been implemented on a wide range of  platforms \cite{Malkova2009,Atala2013,Poli2015,Ling2015}.
Both models possess two bands in their clean incarnation. The SSH model features a chiral symmetry which constraints the Bloch states and allows to define a topological winding number \cite{Ryu2002}. Defects between regions of different winding number introduce  localized, square-normalisable defect states of a fixed chirality that are pinned to the midgap energy. The procedure of taking square roots of lattice systems proposed here provides a mechanism to generated a wider class of models, including models with multiple band gaps, where some of the topological properties  can be traced back to features of a parent system while others emerge from the square-root operation.
Given a suitable parent system with energy bands at positive energies, taking the non-trivial square root provides us with a symmetric arrangement of energy bands at positive and negative energies. If the original system harboured 2$|\nu|$ protected modes around a spectral symmetry point $E_0^2$, the new system will harbour  $|\nu|$  protected modes around energy $E=E_0$ and  $|\nu|$ such modes around energy $E=-E_0$.
Furthermore, the resulting system can also have topologically protected states around the newly emergent spectral symmetry point $E=0$, whose formation is controlled by an independent topological invariant. As we will show, these features arise because the square-root operation allows us to induce (typically nonsymmorphic) \emph{spectral} symmetries at the expense of broken \emph{crystal} symmetries.

\begin{figure}
	\includegraphics[width=\columnwidth]{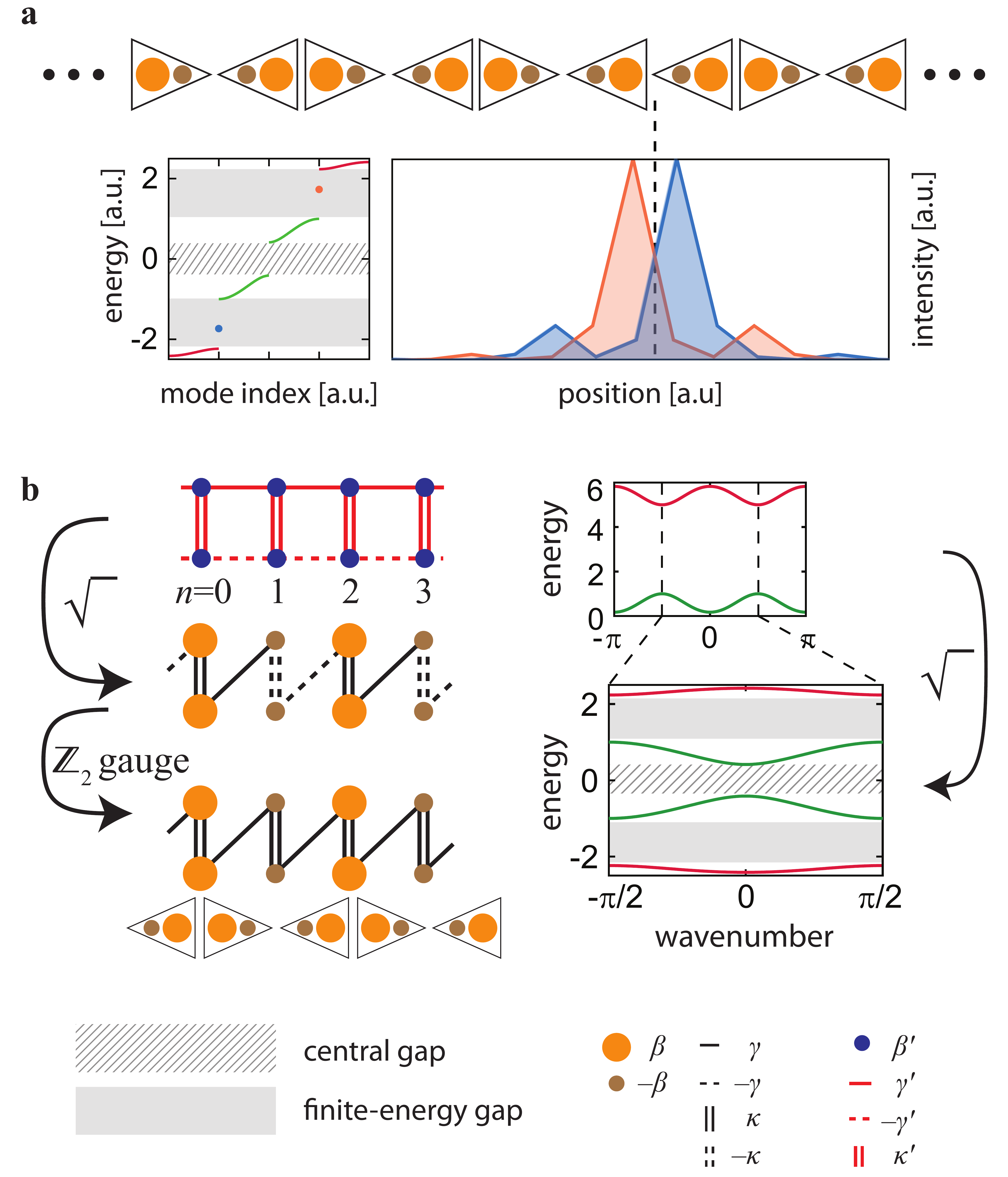}
\caption{\label{fig1} \textbf{Minimal model of a non-trivial square root.}
(a) The bow-tie chain is composed of a sequence of dimers (triangles), each supporting two nondegenerate modes (onsite energies $\pm \beta$ and intradimer coupling $\gamma$) where one mode couples to the left and the other couples to the right (interdimer coupling $\kappa$). In the regular case the dimer orientations alternate, resulting in a periodic system with four bands. The depicted orientation defect generates robust states in both of the finite energy bands. Here this is illustrated for $\beta=\gamma=\kappa=1$, corresponding to the change of the topological index $\tilde\xi$ (see text); further defect configurations are shown in Figs.~\ref{fig4} and \ref{fig5}.
(b)
Interpretation of the dimer chain as a non-trivial square root of a two-legged ladder system (a tight-binding system with $\beta'=\beta^2+\gamma^2+\kappa^2$, $\kappa'=2\beta\kappa$, $\gamma'=\gamma\kappa$). The parent system has two sites per unit cell, thus only features two bands. After taking the non-trivial square root we obtain a tight-binding system with four sites per unit cell, which can be unfolded into a linear chain with nearest-neighbor couplings. The bow-tie chain emerges after a $\mathbb{Z}_2$ gauge transformation, which renders all couplings positive.
}
\end{figure}

\begin{figure}
	\includegraphics[width=\columnwidth]{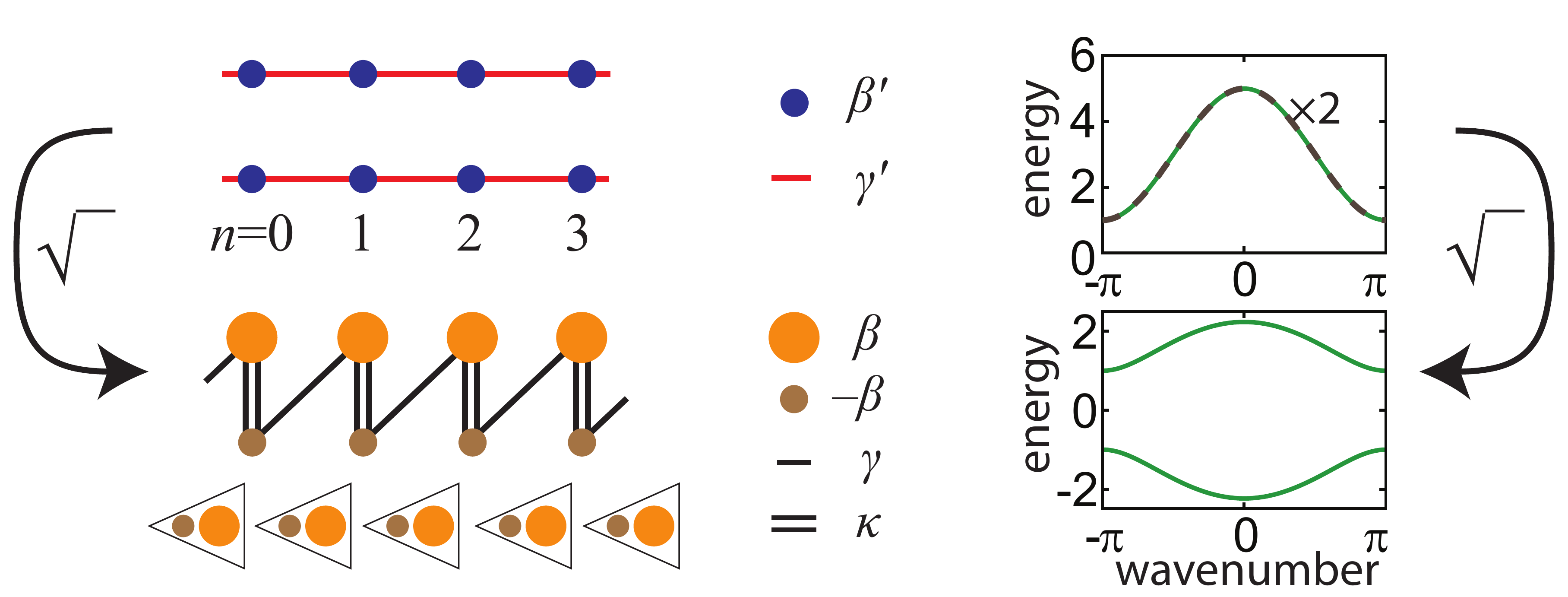}
\caption{\label{fig2} \textbf{Example of a trivial square root.}
The Rice-Mele model, a non-topological system proposed to describe conjugated polymers, is composed of the same dimers as the bow-tie chain (Fig.~\ref{fig1}), but placed in an non-alternating fashion. The Rice-Mele model can also be interpreted as a square root, but possesses the same number of sites per unit cell as its parent system (a system of two uncoupled chains, with effective parameters defined in the same way as in Fig.~\ref{fig1}). The illustrating band structures correspond to $\beta=\gamma=\kappa=1$, hence $\beta'=3$ and $\gamma'=2$.}
\end{figure}

We justify this  proposition with some general preparatory remarks (Section~\ref{sec:prep}), and
then demonstrate the resulting features by deriving a simple minimal model that complies with the properties mentioned above (Section~\ref{sec:bowtie}). Applying a $\mathbb{Z}_2$ gauge transformation, the derived system takes the form of a bow-tie chain (see Fig.~\ref{fig1}), which displays a chiral symmetry involving a fractional lattice translation; the system can also be interpreted as a topological extension of the Rice-Mele model shown in Fig.~\ref{fig2}.
The bow-tie chain allows to explicitly demonstrate the topological nature of the different bands, band gaps and interfaces, as expressed via topological Zak phases, generalized Witten indices, and their mismatch at boundaries and defects (see
Section \ref{sec:top} and
Figs.~\ref{fig3}--\ref{fig5}, as well as the Appendix detailing the utilized scattering approach). To demonstrate that our construction is experimentally accessible and applies to practical devices beyond the tight-binding assumption, we describe the realization of the model in silicon photonics structures, where the $\mathbb{Z}_2$ gauge freedom guarantees that all effective couplings can be made positive (see  Section \ref{sec:phot} and Figs.~\ref{fig6} and \ref{fig7}). Beyond the setting of this paradigmatic model, we then consolidate our general criterion whether a square root of a tight-binding system qualifies as non-trivial --- the resulting system has to exhibit reduced crystal symmetries, manifested e.g. via additional components that give rise to an increased number of bands with newly emerging spectral symmetries --- and identify a number of systems in higher dimensions where this is encountered as well (see Section \ref{sec:general} as well Figs.~\ref{fig8} and \ref{fig9}). As our findings also transfer to analogous realizations in atom-optical and electronic systems (see our concluding Section \ref{sec:conc}), they provide a general route to the design of topologically rich and robust systems with multiple types of protected modes.

\section{Preparatory remarks and premise}
\label{sec:prep}
\subsection{Topological versus non-topological band structures}

To develop our ideas we consider quasi one-dimensional tight-binding systems with nearest-neighbour couplings.
Such systems are defined on a chain of cells labelled by an integer $n$, each associated with an $M$-component amplitude $\boldsymbol{\psi}_n$ (components $\psi_{n,1}\ldots\psi_{n,M}$), where $M$ takes care of the number of sites in the cell, including internal degrees of freedom such as polarization or spin.  The energies $E$ of the system are obtained from the tight-binding equations
\begin{equation}
\label{eq:tb}
E\boldsymbol{\psi}_n=H_n\boldsymbol{\psi}_n+T_{n-1}^\dagger\boldsymbol{\psi}_{n-1}+T_{n}\boldsymbol{\psi}_{n+1}
\end{equation}
with on-site Hamiltonians $H_{n}=H_{n}^\dagger$ and nearest-neighbor coupling matrices $T_{n}$. A band structure emerges when the system is periodic. We then can write the eigenstates as Bloch waves $\boldsymbol{\psi}_{n}=\boldsymbol{\varphi}(k)e^{ink}$ with wave number $k$, and find the $k$-dependent eigenvalue problem $E(k)\boldsymbol{\varphi}(k)=H(k)\boldsymbol{\varphi}(k)$ with
Bloch Hamiltonian $H(k)=H_0+T_{0}^\dagger e^{-ik}+T_{0}e^{ik}$. The number $M$ of sites in the unit cell determines the dimensions of the Bloch Hamiltonian, and thereby also the number of energy bands $E(k)$, associated with eigenvectors $\boldsymbol{\varphi}(k)$.

A useful prepatory example is the
the Rice-Mele model (shown in Fig.~\ref{fig2}),  where $M=2$ and
\begin{equation}
H_n
=\left(\begin{array}{cc} \beta & \kappa \\ \kappa & -\beta\end{array}\right),\quad T_n
=\left(\begin{array}{cc} 0 & 0 \\ \gamma & 0\end{array}\right).
\end{equation}
It will be convenient to represent this system as a sequence of asymmetric dimers with  onsite energies $-\beta$, $\beta$, intradimer coupling $\gamma$ and interdimer coupling $\kappa$; these dimers are depicted symbolically as triangles in all figures.
(Note that these dimers combine two sites of adjacent cells, i.e., an amplitude $\psi_{n,2}$ with an amplitude $\psi_{n+1,1}$; see the Appendix and Fig. \ref{figapp} for further details.)
The two bands $E(k)=\pm\sqrt{\beta^2+|\kappa+\gamma e^{ik}|^2}$ are arranged symmetrically in energy, which can be associated with the property
$H(k)=-\sigma_y [H(k)]^*\sigma_y$, where  $\sigma_l$ denotes the standard Pauli matrices. This band structure is not classified as topological
as the line $\beta=0$, $|\gamma|=|\kappa|$, where the gap closes, does not divide the parameter space $(\beta,\gamma,\kappa)$  into disconnected segments. For fixed $\beta=0$, however, the system reduces to Su-Schrieffer Heeger model, whose band structure is topological as the gap-closing lines $|\kappa|=|\gamma|$ now indeed divide the reduced parameter space  $(\gamma,\kappa)$ into disconnected segments. The passage across a gap-closing line is known as a band inversion. The topological properties of the Su-Schrieffer-Heeger model can be associated with the chiral symmetry $\sigma_z H(k) \sigma_z=-H(k)$ \cite{Ryu2002}, which lifts the system into a topologically non-trivial universality class. It is also useful to note that the chiral operator $\sigma_z$ constitutes a special case of a $\mathbb{Z}_2$ gauge transformation, in general given by $\mathcal{Z}\psi_{n,m}=\sigma_{nm} \psi_{n,m}$ with independently chosen $\sigma_{nm}=\pm 1$. Such a gauge transformation allows to change the sign of off-diagonal elements (couplings) in a tight-binding Hamiltonian.

\subsection{General considerations}
\label{sec:prop}
For further motivation, let us  have a more general look at periodic tight-binding systems with Bloch Hamiltonian $H(k)$, which may include an arbitrary range of the couplings, and explore a particular consequence of a chiral symmetry $H(k)\mathcal{X}(k)=-\mathcal{X}(k)H(k)$. It will be consistent with the following discussions to set the dimensions of this Hamiltonian to $2M$, as it is our goal to relate it to a parent Hamiltonian with only $M$ components.
As indicated, we acknowledge that the chiral operator $\mathcal{X}(k)$ may be $k$ dependent. In a suitable, possibly  $k$-dependent basis (viz., gauge), however, we can fix $\mathcal{X}(k)=\sigma_x\otimes \openone_{M}\equiv \mathcal{X}$ \cite{remark_flat_bands}, upon which the Bloch Hamiltonian takes the form
\begin{equation}
H(k)=\left(\begin{array}{cc}U(k) & V(k) \\ -V(k) & -U(k)
\end{array}
\right), \quad \begin{cases} U^\dagger(k)=U(k), \\ V^\dagger(k)=-V(k). \end{cases}\!\!\!
\end{equation}
For any eigenvector $\boldsymbol{\varphi}(k)=(\mathbf{u}(k), \mathbf{v}(k))^T$ with energy $E(k)$, we have a partner state $\mathcal{X}\boldsymbol{\varphi}(k)=(\mathbf{v}(k), \mathbf{u}(k))^T$ with energy $-E(k)$; also, $\mathbf{u}^\dagger(k) \mathbf{v}(k)+\mathbf{v}^\dagger(k) \mathbf{u}(k)=0$.

Note now that the squared Hamiltonian
\begin{equation}
H^2(k)=\left(\begin{array}{cc}U^2(k)-V^2(k) & U(k)V(k)-V(k)U(k) \\ U(k)V(k)-V(k)U(k) & U^2(k)-V^2(k) \end{array}\right)
\end{equation}
commutes with $\mathcal{X}(k)$. The joint eigenvectors are the superpositions
\begin{equation}
(\openone\pm \mathcal{X}) \boldsymbol{\varphi}(k)=\left(\begin{array}{c}\mathbf{u}(k)\pm \mathbf{v}(k) \\ \mathbf{v}(k)\pm \mathbf{u}(k)\end{array}\right) =\left(\begin{array}{c}\mathbf{u}'_\pm(k) \\ \pm \mathbf{u}'_\pm(k)\end{array}\right),
\end{equation}
where $E^2(k) \mathbf{u}'_\pm(k)=H'_{\pm}(k)\mathbf{u}'_\pm(k)$ with the reduced Hamiltonian $H'_{\pm}(k)=(U(k)\mp V(k))(U(k)\pm V(k))$.

Normally, we would expect $U(k+nk_0)=U(k)$ and $V(k+nk_0)=V(k)$ be both periodic when cycling through the Brillouin zone of size $k_0$. However, as we will confirm in our concrete example, the gauge choice that renders $\mathcal{X}(k)$ constant can render
$V(k)$ antiperiodic, $V(k+nk_0)=(-1)^n V(k)$. In this case, traversing the Brillouin zone joins $H'_-(k)=H'_+(k+k_0)$ and $\mathbf{u}'_-(k)=\mathbf{u}'_+(k+k_0)$, so that it is natural to double the size of the Brillouin zone and describe the Bloch waves by a reduced set of $M$ (instead of $2M$) components. Furthermore, applying the transformation $\boldsymbol{\varphi}(k)\to \mathrm{diag}\,(e^{i\pi k/2k_0},e^{-i\pi k/2k_0})\boldsymbol{\varphi}(k)$ we can revert to a gauge where $H(k)$ is periodic, and then find that $\mathcal{X}(k)=[\cos(\pi k/k_0)\sigma_x+\sin(\pi k/k_0)\sigma_y]\otimes \openone_{M}$ takes the form of a fractional lattice translation, i.e., can be interpreted as a nonsymmorphic chiral symmetry.

Upon retracing our steps, the upshot of this discussion is the following proposition: taking a square root of a parent tight-binding system as described by $H'_+(k)$, it can be possible to  break crystal symmetries, at the expense of an expanded unit cell with twice as many components and a Brillouin zone half in size, and in the process generate a chiral symmetry $\mathcal{X}$ that induces spectrally symmetric bands. We will show that this can indeed be achieved, including in cases where the band structure is already topological, resulting in a non-trivial square-root system which displays richer topological features. In particular, we construct a practically realisable model system that only features nearest-neighbour couplings.

\section{The bow-tie chain}
\label{sec:bowtie}
\subsection{Construction of the minimal model}
\label{sec:construc}
According to the features described in the preceding Section, we require that our minimal parent system features a band gap about a spectral symmetry point; the unit cell therefore needs to comprise at least two sites ($M=2$). To be non-trivial, the square-root system will have to have four bands, thus, be periodic with period 2 in the original unit-cell indices. Together with the required spectral symmetries, this allows to identify a minimal model, which will turn out to correspond to the bow-tie chain depicted in Fig. \ref{fig1}.

To implement these constraints we start with the putative non-trivial square root system (termed the `candidate') and iterate the tight-binding equations \eqref{eq:tb} once, giving
\begin{equation}
\label{eq:iter1}
E'\boldsymbol{\psi}_n=H_n'\boldsymbol{\psi}_n+T_{n-1}'^\dagger\boldsymbol{\psi}_{n-1}+T_{n}'\boldsymbol{\psi}_{n+1}+{\tilde T}_{n-2}^{\prime\dagger}\boldsymbol{\psi}_{n-2}+{\tilde T}_{n}'\boldsymbol{\psi}_{n+2}
\end{equation}
where
\begin{subequations}
\begin{align}
E'&=E^2,\\
H_n'&=H_n^2+T_{n-1}^\dagger T_{n-1}+T_{n} T_{n}^\dagger,
\\
T_{n}'&=H_nT_{n} + T_nH_{n+1},
\\
{\tilde T}_{n}'&= T_{n}T_{n+1}.
\end{align}
\label{eq:iter2}%
\end{subequations}
We interpret this as a new tight-binding system, describing the parent system with a positive energy spectrum.
This parent system should be periodic,
$H_n'=H_0'$, $T_{n}'=T_0'$, with vanishing next-nearest neighbour couplings ${\tilde T}_{n}'=0$.
As shown in Fig.~\ref{fig2}, taking  the Rice-Mele model as the candidate this construction leads to a parent system consisting of two uncoupled chains with onsite energies $\beta'=\beta^2+\gamma^2+\kappa^2$ and couplings $\gamma'=\gamma\kappa$; the Rice-Mele model then constitutes a trivial square root without any newly emerging topological features, consistent with the fact that it still has the same period as its parent.

To ensure that we obtain a non-trivial square root, with four bands arranged symmetrically about $E=0$, we demand that the candidate system has a period of two, and possesses a chiral symmetry which maps $E$ to $-E$. This can be enforced by the choice
\begin{equation}
\label{eq:chiraltb}
 H_n=H_0 (-1)^n,\quad T_n=T_0 (-1)^n.
\end{equation}
The chiral symmetry is induced by a translation $\mathcal{X}\boldsymbol{\psi}_n= \boldsymbol{\psi}_{n+1}$, thus, a fractional translation by half a period of the candidate system. This spectral symmetry is trivial in the parent system, and therefore emerges only upon taking the square root, at the expense of a reduced translational crystal symmetry. The constraint ${\tilde T}_{n}'=0$ requires $T_0^2=0$. The freedom to choose the basis in every cell then allows us to write
\begin{equation}
H_n=(-1)^n\left(\begin{array}{cc} \beta & \kappa \\ \kappa & \beta \end{array}\right),\quad
T_n=(-1)^n\left(\begin{array}{cc} 0 & 0 \\ \gamma & 0 \end{array}\right),
\end{equation}
which defines the minimal non-trivial square root system.
Its parent system is given by
\begin{subequations}
\begin{align}
H_n'&=
\left(\begin{array}{cc} \beta^2+\gamma^2+\kappa^2 & 2\beta\kappa \\ 2\beta \kappa &  \beta^2+\gamma^2+\kappa^2 \end{array}\right)
,
\\
T_{n}'&=
\left(\begin{array}{cc} \gamma\kappa & 0 \\ 0 & -\gamma\kappa \end{array}\right)
.
\end{align}%
\end{subequations}
Via a suitable $\mathbb{Z}_2$ gauge transformation
we can enforce that the couplings $\gamma,\kappa\geq 0$ are real and nonnegative, which we will assume from hereon.

\subsection{Interpretation and band structure}
\label{sec:interpretation}

With help of Fig.~\ref{fig1}(b), we now can  confirm that the parent Hamiltonian corresponds to a two-legged ladder, with onsite energies $\beta'=\beta^2+\gamma^2+\kappa^2$, couplings $\gamma'=\gamma\kappa$ and $-\gamma'$ along the two legs and coupling $\kappa'=2\beta \kappa$ along the rungs.
Given the opposite couplings along the legs, each plaquette is penetrated by a flux phase of $\pi$. The Bloch Hamiltonian of the parent is
\begin{equation}
H'(k)=\left(\begin{array}{cc} \beta' + 2\gamma'\cos k &\kappa'\\ \kappa' & \beta' - 2\gamma'\cos k \end{array}\right),
\end{equation}
and the two energy bands are
\begin{equation}
E'_{\mu}(k)=\beta'+\mu\sqrt{\kappa'^2+4\gamma'^2\cos^2 k}, \quad\mu=\pm1.
\end{equation}

As depicted in the figure, the non-trivial square root system can be unfolded into a linear chain, where all couplings are still restricted to nearest neighbours (see the Appendix for a very detailed description). Along the chain, the system then displays a repeating coupling sequence $\kappa$, $\gamma$, $-\kappa$, $-\gamma$, and a repeating sequence of onsite energies $\beta$, $\beta$, $-\beta$, $-\beta$.
We interpret this as an extended Rice-Mele model, but with topological features that we establish in the next Section. For the photonic implementation we will make all couplings  positive by an additional $\mathbb{Z}_2$ gauge transformation; schematically, the system is then composed of oppositely orientated dimers that form the  repeating bow-tie pattern shown in Fig.~\ref{fig1}. For the analytical considerations, it is more convenient to retain the system with the coupling sequence as derived.

The Bloch Hamiltonian of the unfolded linear chain is
\begin{equation}
\label{eq:bloch}
H(k)=\left(\begin{array}{cccc} \beta &\kappa & 0 & -\gamma e^{-ik} \\ \kappa & \beta & \gamma e^{ik} & 0
\\
0 & \gamma e^{-ik} & -\beta  & -\kappa\\ -\gamma e^{ik} & 0 &-\kappa & -\beta
 \end{array}\right),
\end{equation}
and the four energy bands are
\begin{align}
\label{eq:bands}
E_{\mu,\eta}(k)&=\eta\sqrt{E'_{\mu}(k)} \nonumber \\
&=\eta\sqrt{\beta^2+\gamma^2+\kappa^2+2\mu\kappa
\sqrt{\beta^2+\gamma^2\cos^2k}},
\end{align}
where the label $\eta=\pm 1$ selects the bands at positive and negative energies.
The four bands
thus come in two pairs,
covering the ranges
\begin{subequations}
\begin{align}
&\xi\varepsilon_-<|E|<\tilde\varepsilon_{-\tilde\xi}&\quad\mbox{(inner bands, $\mu=-1$)},\\
&\tilde\varepsilon_{\tilde\xi}<|E|<\varepsilon_+&\quad\mbox{(outer bands, $\mu=1$)},
\end{align}
\label{eq:edges}%
\end{subequations}
separated by gaps at
\begin{subequations}
\begin{align}
&|E|<\xi\varepsilon_-&\quad\mbox{(central gap)},\\
&\tilde\varepsilon_{-\tilde\xi}<|E|<\tilde\varepsilon_{\tilde\xi}&\quad\mbox{(finite-energy gaps)}.
\end{align}%
\end{subequations}
The band edges are given by
\begin{subequations}
\begin{align}
\varepsilon_\pm&=\sqrt{\beta^2+\gamma^2}\pm\kappa,\\
\tilde\varepsilon_\pm&=\sqrt{\gamma^2+(\beta\pm\kappa)^2}.
\end{align}%
\end{subequations}
These expressions feature an index
\begin{align}
\xi&=\mathrm{sgn}\,\varepsilon_-=\mathrm{sgn}\,(\beta^2+\gamma^2-\kappa^2),
\end{align}
which changes its sign when the central gap closes while parameters are steered through a band inversion of the inner bands. Analogously, the index
\begin{align}
\tilde\xi&=\mathrm{sgn}\,(\tilde\varepsilon_+-\tilde\varepsilon_-)=\mathrm{sgn}\,\beta
\end{align}
changes its sign in a band inversion in which the finite-energy gaps close ($\beta=0$, where we recover the Su-Schrieffer-Heeger model with only a single, central, gap).

\begin{figure*}
	\includegraphics[width=1.6\columnwidth]{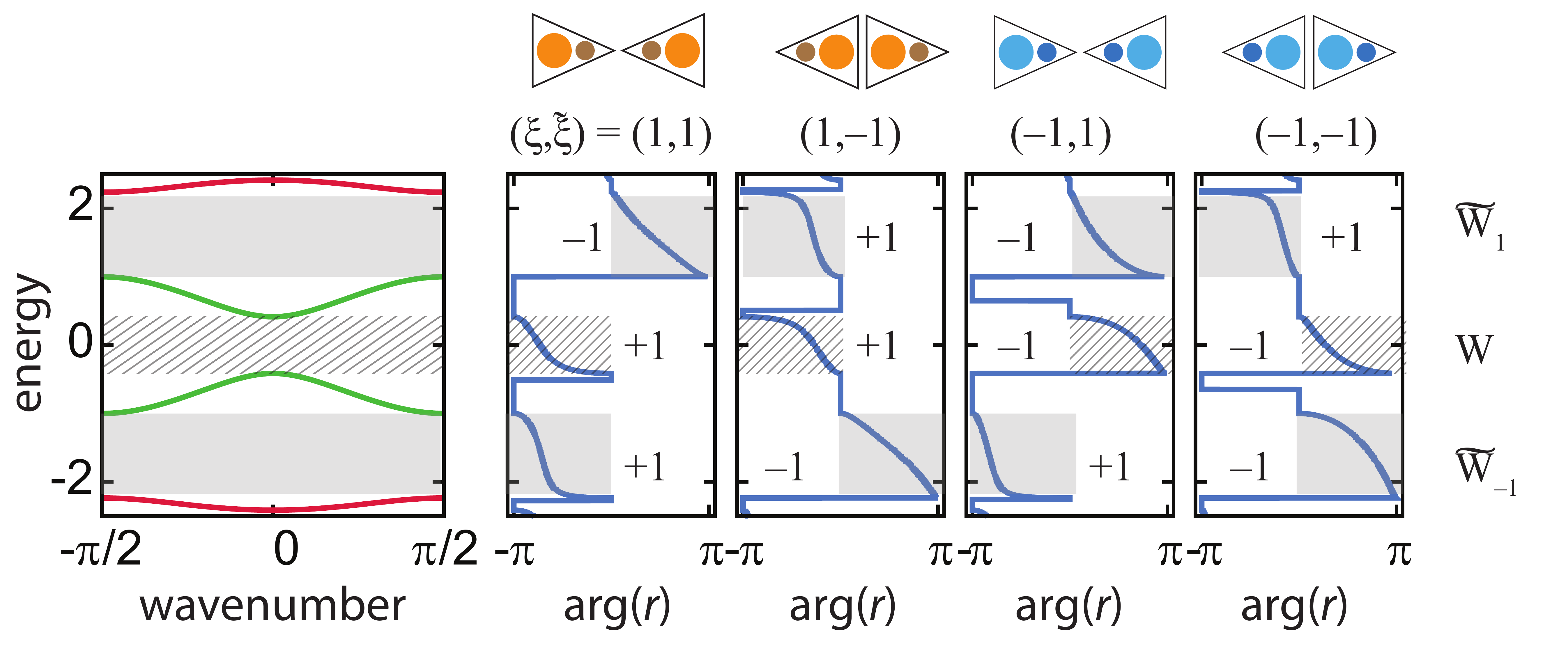}
\caption{\label{fig3} \textbf{Witten index in different configurations.}
The left panel shows the identical band structure for four different parameter combinations, corresponding to the four combinations of the topological index $\xi=\pm1$, $\tilde \xi=\pm1$. The remaining panels show the winding of the reflection phase, obtained from the reflection coefficient \eqref{eq:refl}, and the Witten index in each of the bands.
For $(\xi,\tilde  \xi)=(1,1)$ we set $\beta=\gamma=\kappa=1$, while for
$(\xi,\tilde  \xi)=(1,-1)$ we set $-\beta=\gamma=\kappa=1$, thus change the sign of $\beta$ (equivalently, interchange the orientation of the dimers, as shown on the top).
For $(\xi,\tilde  \xi)=(-1,1)$ we transform the parameters according to Eq.~\eqref{eq:partrafo}, resulting in $2\beta=2\gamma=\kappa=\sqrt{2}$ (the dimers corresponding to these transformed values are shown in blue). The case
$(\xi,\tilde  \xi)=(-1,-1)$ follows by once more changing the value of $\beta$,  so that $-2\beta=2\gamma=\kappa=\sqrt{2}$; this is again equivalent to interchanging the orientation of the dimers.
}
\end{figure*}

The indices $\xi$ and $\tilde \xi$ encode information which cannot be inferred by simply inspecting the band structure --- the same band structure is found when one changes the sign of $\beta$, which changes the sign of $\tilde\xi$; the band structure is also invariant when one passes over to parameters
\begin{equation} \bar\kappa=\sqrt{\beta^2+\gamma^2},\quad \bar\beta=\beta\kappa/\bar\kappa,\quad \bar\gamma=\gamma\kappa/\bar\kappa,
\label{eq:partrafo}
\end{equation}
which changes the sign of $\xi$. In Figs.~\ref{fig3}-\ref{fig5}, the four cases delivering the same band structure are distinguished via the orientation of the dimers (which effectively controls the sign of $\beta$ and thus $\tilde \xi$), while dimers with $\xi=1$ are denoted in orange-brown, and dimers with $\xi=-1$ are denoted in blue.
As we show next, the indices $\xi$ and $\tilde \xi$ indeed capture the topological features of the band structure.

\section{Topological characterization}
\label{sec:top}

To establish the topological features of the bow-tie chain, we identify its symmetries and describe how they relate to the associated topological indices.
The index $\xi$ arises from the chiral symmetry that emerges by taking the square root, while the index $\tilde \xi$ is inherited from the chiral symmetry of the parent system and translates into an uncommon algebraic property of the linear chain. These indices can be expressed in terms of winding numbers in the bands and in the gaps, while interfaces between regions with different indices give rise to topologically protected defect states.

\subsection{Symmetries}
\label{sec:symmetries}

The spectral symmetry $E_{\mu,-}(k)=-E_{\mu,+}(k)$
of the band structure \eqref{eq:bands}
about $E=0$ is a consequence of the chiral symmetry \eqref{eq:chiraltb}, which for the Bloch Hamiltonian
\eqref{eq:bloch} takes the form
\begin{equation}
\mathcal{X}H(k)\mathcal{X}=-H(k), \quad \mathcal{X}=
\left(\begin{array}{cccc} 0 &0 & 1 &0 \\ 0 & 0 & 0 & 1
\\
1 &0 & 0 & 0\\ 0 & 1 &0 & 0
\end{array}\right).
\end{equation}
As mentioned before, this symmetry originates from a fractional lattice translation, by half a period of the system.
We also note the relations
\begin{equation}
\label{eq:symm}
\mathcal{R}H(k)\mathcal{R}=H(-k)=H^*(k)
, \quad \mathcal{R}=\left(\begin{array}{cccc} 0 & 1 &0 &0 \\ 1 & 0 & 0 & 0
\\
0 &0 & 0 & -1\\ 0 & 0 &-1 & 0
\end{array}\right),
\end{equation}
which correspond to a reflection symmetry and a conventional time-reversal symmetry.
Both entail that the bands also are symmetric in $k$, which simplifies the determination of topological indices \cite{Xiao2014,Shi2016}.

The spectral symmetry $E^2_{-,\eta}(k)
=2\beta'-E^2_{+,\eta}(k)$ of the squared bands about $E^2=\beta'$ is the consequence of
the chiral symmetry
$\sigma_yH'(k)\sigma_y=2\beta'-H'(k)$ in the parent system.
For the unfolded linear chain, this gives the remarkable algebraic relation
\begin{equation}
\label{eq:nonlinrelation}
\tilde{\mathcal{X}}H^2(k)\tilde{\mathcal{X}}=2\beta'-H^2(k), \quad \tilde{\mathcal{X}}=
\left(\begin{array}{cccc} 1 & 0 &0 &0 \\ 0 & -1 & 0 & 0
\\
0 &0 & -1 & 0\\ 0 & 0 &0 & 1
\end{array}\right),
\end{equation}
a property which would be difficult to interpret  without knowledge of the underlying parent system. This inherited spectral symmetry plays an important role throughout the remainder of this work.

Given these symmetries, applying the conventional classification of free-fermion models we  expect that the topological features in the central gap can be captured by the symmetry class BDI for chiral systems with a conventional time-reversal symmetry \cite{Ryu2010}, i.e., the same symmetry class as for the SSH model. As explained in the Appendix by adapting the considerations in Ref. \cite{Shiozaki2014}, the same expectation is born out when we take into account that the chiral symmetry encountered in our model is nonsymmorphic. For the finite-energy gaps, we arrive at the same conclusions starting from the topological features of the parent system, which also suggests that the corresponding topological index is independent. To show that these expectations indeed hold true we explicitly construct the topological invariants for the different bands and band gaps.

\subsection{Zak phase and Witten index}
\label{sec:indices}

Each band can be associated with a Zak phase \cite{Zak1989}
\begin{equation}
\label{eq:zak}
z=i\oint_{\mathrm{BZ}}\boldsymbol{\varphi}^\dagger(k)\frac{d}{dk}\boldsymbol{\varphi}(k)=\pi Z.
\end{equation}
This phase depends on a gauge choice, which can be fixed by demanding that the component $\varphi_1(k)$ is real and positive.
In topological systems, the phase is quantized, giving rise to an integer index $Z=z/\pi$, while in non-topological systems the phase can take any value \cite{Ryu2002}.

To evaluate the Zak phase we adopt the convenient scattering formalism (see Refs.~\cite{Meidan2011,Fulga2012,Xiao2014,Poshakinskiy2015,Shi2016}, as well as the Appendix providing further details for the statements in the present Subsection). The phase is then expressed in terms of a reflection coefficient $r(E)$ of a wave entering a semi-infinite segment of the system.
In the band gaps, including the band edges, the wave will be totally reflected, so that $|r(E)|$=1.
By inspecting the winding deep in the bulk of each different topological sector, we find the remarkably simple relation
\begin{equation}
\label{eq:zak2}
Z=\frac{1}{2}[r(\mbox{lower band edge})-r(\mbox{upper band edge})]
\end{equation}
for the Zak phase in each band.

The reflection coefficient can be obtained from the transfer matrix of the system. Given the amplitudes $\boldsymbol{\psi}_n$ in a cell, the tight-binding equations at fixed energy $E$ allow to determine the amplitudes in the next cell as $\boldsymbol{\psi}_{n}=M_n(E)\boldsymbol{\psi}_{n-1}$. The matrix $M(E)=M_2(E)M_1(E)$ then describes the transfer by two cells, thus a period of the square-root system. Unitarity of quantum mechanics enforces the symplectic symmetry $M^\dagger(E) \sigma_y M(E)=\sigma_y$, which amounts to flux conservation. This entails ${\rm det}\,M(E)=1$, so that the two eigenvalues $\Lambda_\pm(E)$ of $M(E)$ are reciprocal, $\Lambda_-(E)=1/\Lambda_+(E)$. In the bands, these eigenvalues determine the Bloch factors of the propagating waves. In the gap, the modes become evanescent, where we choose $|\Lambda_+(E)|<1$ to describe the decaying wave.
The reflection coefficient
\begin{equation}
r(E)=\frac{\phi_{+,1}(E)+i\phi_{+,2}(E)}{\phi_{+,2}(E)+i\phi_{+,1}(E)}
\label{eq:refl}
\end{equation}
follows from the associated eigenvector $\boldsymbol{\phi}_+(E)$.

At the band edges, the propagating modes must match up with the evanescent modes. This enforces $\Lambda_+=\Lambda_-=1$ or $\Lambda_+=\Lambda_-=-1$, as well as $\boldsymbol{\phi}_+=\boldsymbol{\phi}_-=(1,1)^T$ or $\boldsymbol{\phi}_+=\boldsymbol{\phi}_-=(1,-1)^T$, corresponding to $r=1$ or $r=-1$.
At the band edges $E=\pm\varepsilon_{-}$ of the central gap, we find $\Lambda_+=1$, $\boldsymbol{\phi}_+=(1,\mp 1)^T$, thus $r=\mp 1$.
At the band edges $|E|=\tilde\varepsilon_{\pm}$ of the finite-energy gap, we have  $\Lambda_+=-1$ and
$\boldsymbol{\phi}_+=(1,\pm 1)^T$, thus $r=\pm 1$.  At the extremal edges $E=\pm\varepsilon_{+}$, we have $\Lambda_+=1$ and $\boldsymbol{\phi}_+=(1,\pm 1)^T$, so that $r=\pm 1$. The Zak phase of each band is therefore indeed quantized, and can be written as
\begin{equation}
\label{eq:zakresult}
Z_{\mu,\eta}=\left\{
               \begin{array}{ll}
                 (\eta\tilde\xi-\xi)/2, & \mbox{(inner bands, $\mu=-1$)},\\
                 (\eta\tilde\xi-1)/2, & \mbox{(outer bands, $\mu=1$)}. \\
               \end{array}
             \right.
\end{equation}
In particular, we can express $\xi=-(Z_{-1,1}+Z_{-1,-1})$ and
$\tilde\xi=(Z_{-1,1}-Z_{-1,-1})=(Z_{1,1}-Z_{1,-1})$.

Similarly, we can associate a topological phase to each gap. This can normally done, e.g., via the Witten index \cite{Witten1982}, which here relates to the reflection phase at a spectral symmetry point \cite{Niemi1986,Bolle1987,Borisov1988}. In the finite-frequency gaps, this information is instead encoded in the winding of the reflection coefficient $r(E)=\exp(i\phi(E))$ as one crosses the gap from the lower band edge to the upper band edge (see Fig.~\ref{fig3}). At the band edges, the phase $\phi(E)$ is fixed to the symmetry-protected values $0$ or $\pi$, so that $r=\pm1$ as discussed above. As one crosses the gap, the reflection coefficient winds along the unit circle, which encodes topological information.
As dictated by causality \cite{deCarvalho2002}, the winding is always in the clockwise sense, and in the system considered here, is always by $\pi$. A reflection coefficient starting at $r=1$ at the lower band edge therefore need to pass by the point $r=-i$  before ending at $r=-1$ at the upper band edge; this scenario is characterized by the Witten index $1$. If the arc is from $r=-1$ via $r=i$ to $r=1$, we associate this with an index $-1$. With these conventions it follows that the Witten index is equal to the value of $r$ at the lower band edge,
\begin{subequations}
\begin{align}
W=\xi \quad\mbox{(central gap)},\\
\tilde W_\eta=-\eta\tilde\xi\quad\mbox{(finite-energy gaps)}.
\end{align}
\label{eq:wittenres}%
\end{subequations}
In each gap, the Witten index is therefore directly related to the index $\xi$ or $\tilde\xi$ that controls the band inversion.
This establishes the topological features of the bandstructure. We now turn to the observable consequences, and in particular the formation of defect states.

\begin{figure}
	\includegraphics[width=\columnwidth]{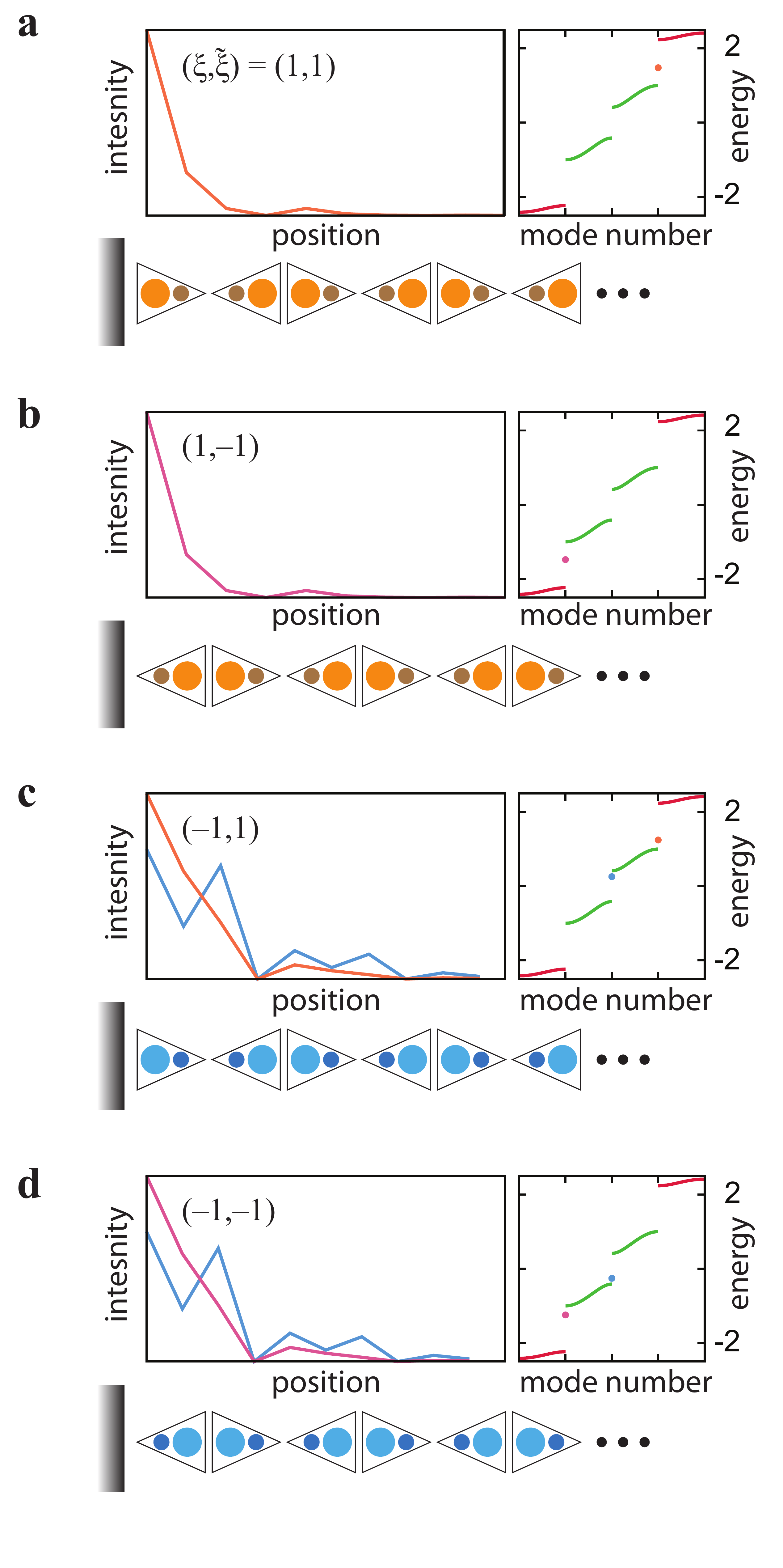}
\caption{\label{fig4} \textbf{Edge states.}
Each figure (a-d) corresponds to a semi-infinite system with parameters as given in Fig.~\ref{fig3}. A defect state exists whenever the Witten index in a gap equals $=1$.
In each figure, the left panel shows the mode profile of the edge state while the right panel indicates its  position within the spectrum of the system.
}
\end{figure}

\begin{figure}
	\includegraphics[width=\columnwidth]{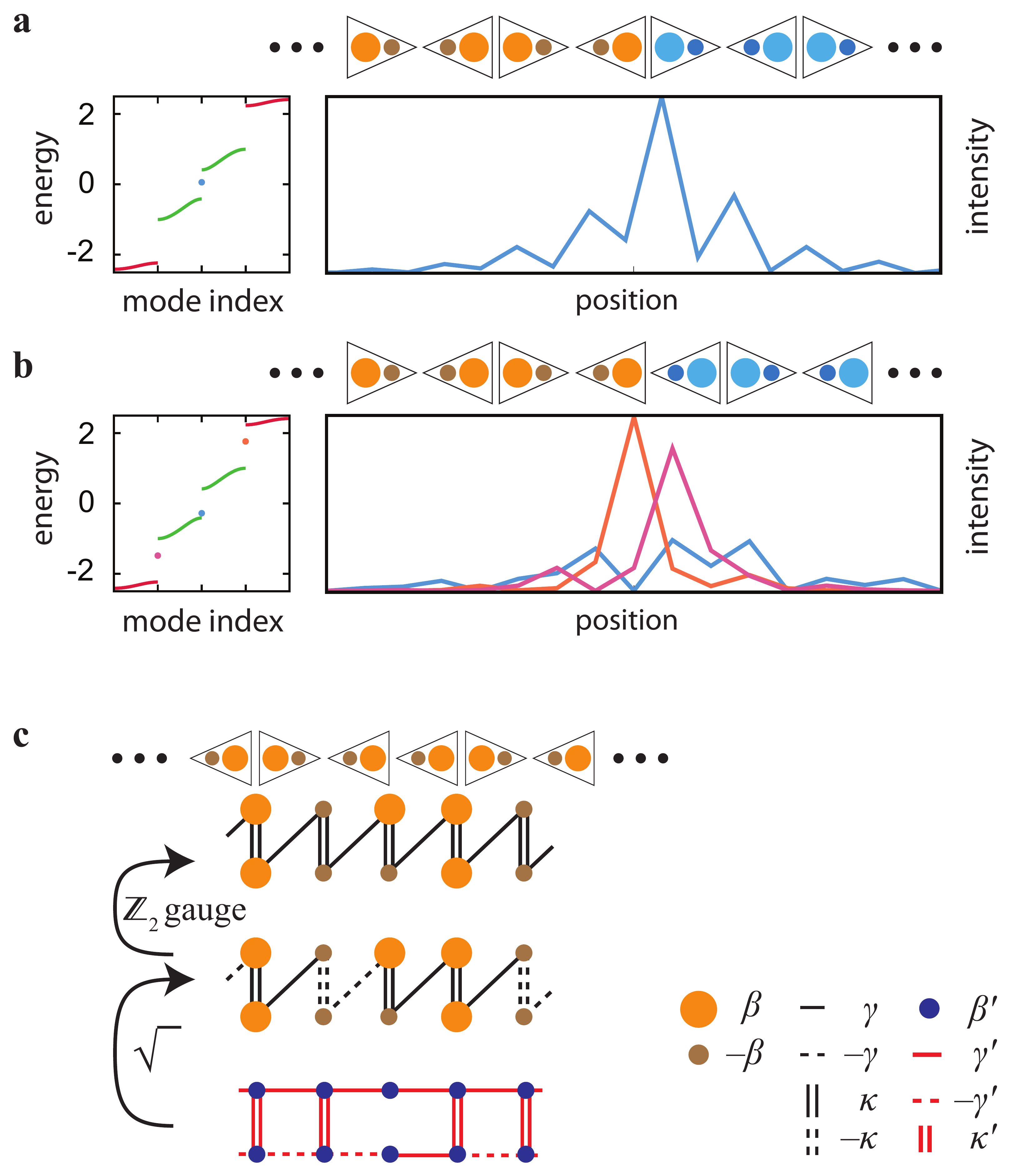}
\caption{\label{fig5} \textbf{Interface states.}
Defect states that arise at interfaces between systems with opposite index $\xi_L=1$,  $\xi_R=-1$, corresponding to a change of the parameters by the transformation \eqref{eq:partrafo}. In panel (a), the index $\tilde \xi_L=\tilde \xi_R=1$ is identical on both sides, while in panel (b) we have $\tilde\xi_L=1$, $\tilde \xi_R=-1$.  The parameters are chosen as in Figs.~\ref{fig3} and \ref{fig4}.
See Fig.~\ref{fig1}(a) for an interface state between systems with opposite indices $\tilde \xi_L=1$, $\tilde \xi_L=-1$ but identical $\xi_L=\xi_R=1$; as shown in panel (c), this interface amounts to a coupling defect in the parent system. No such interface states form when all indices are the same on both sides of an interface.
}
\end{figure}

\subsection{Topologically protected defect states}
\label{sec:topstates}
The adopted scattering approach to the topological characterization of the bandstructure sets up an efficient criterion to infer the existence of topologically protected defect states.

When we terminate the system as shown in Fig.~\ref{fig4}, a bound state is formed if the decaying state fulfills the boundary condition
$\phi_{+,1}(E)=0$, hence $r(E)=i$, which is guaranteed for a Witten index $W=-1$ in the central gap, and $\tilde W_\eta=-1$ in the finite-energy gaps.
Since the Witten index in the finite energy gap at positive energies is opposite to the one at negative energies, exactly one such finite-energy bound state exists for any given combination of parameters; in the central gap, the state only exists for
some configurations. In particular, the state in the central gap can be switched on or off by passing from the values $(\beta,\gamma,\kappa)$  to the transformed values
$(\bar\beta,\bar\gamma,\bar\kappa)$ given in Eq.~\eqref{eq:partrafo}, which changes the index $W$ while keeping the bulk bandstructure unchanged. In the Figure, dimers with $W=1$ are again denoted in orange-brown, while the transformed dimers with $W=-1$ are denoted in blue.

These considerations can be extended to more general boundary conditions. E.g., a boundary cutting through the middle of a dimer translates into the condition $\phi_{+,2}(E)=0$, hence $r(E)=-i$, requiring a Witten index $W=1$ in the central gap and $\tilde W_\eta=1$ in the finite-energy gaps. Displacing the boundary by a full dimer amounts to the fractional lattice translation that induces the chiral symmetry \eqref{eq:chiraltb}. Under this translation, the index $\tilde W$ (and hence also $\tilde \xi$) changes its sign while $W$ (and hence also $\xi$) remains unchanged. This is consistent with the results shown in Fig.~\ref{fig4}, where such a translation interchanges panels (a) and (b), as well as panels (c) and (d).

As already anticipated in Fig.~\ref{fig1}(a), defect states can also form at interfaces  between two semi-infinite systems, denoted as left (L) and right (R). In the parent Hamiltonian, this defect amounts to a coupling defect between to the two ladders, as shown in Fig.~\ref{fig5}(c).
Two further interface scenarios are shown in Fig.~\ref{fig5}(a,b). In terms of the corresponding reflection matrices, the general quantization condition of states in the gaps can then be written as $R_{LR}\equiv r_Lr_R=1$.
As we traverse through a gap, this product will rotate by $2\pi$ along the unit circle, and  is guaranteed to pass through $1$ if at the band edges $R_{LR}=-1$. This occurs exactly when the Witten index of the gap differs on the two sides of the interface. The central gap supports a topological defect state if $\xi_L\neq \xi_R$, while a topological defect state in both of the finite-frequency gaps appears if $\tilde\xi_L\neq \tilde\xi_R$. We recall that all combinations of these indices can be achieved while keeping the bandstructure on both sides aligned. In particular, to set $\xi_L\neq \xi_R$, one can again pass over to the values from values $(\beta,\gamma,\kappa)$ on one side to the transformed values $(\bar\beta,\bar\gamma,\bar\kappa)$ on the other side; the different types of dimers are again indicated by their color.

Therefore, by either using a boundary or an interface, the formation of topological protected bound states can be controlled independently in any of the three gaps.
In contrast to the SSH model, these states do not display a sublattice polarization; however, they all decay exponentially and therefore are square normalizable.
We next demonstrate the utility of these states in the specific setting of silicon photonics.

\section{Photonic realization}
\label{sec:phot}

\begin{figure}
	\includegraphics[width=0.8\columnwidth]{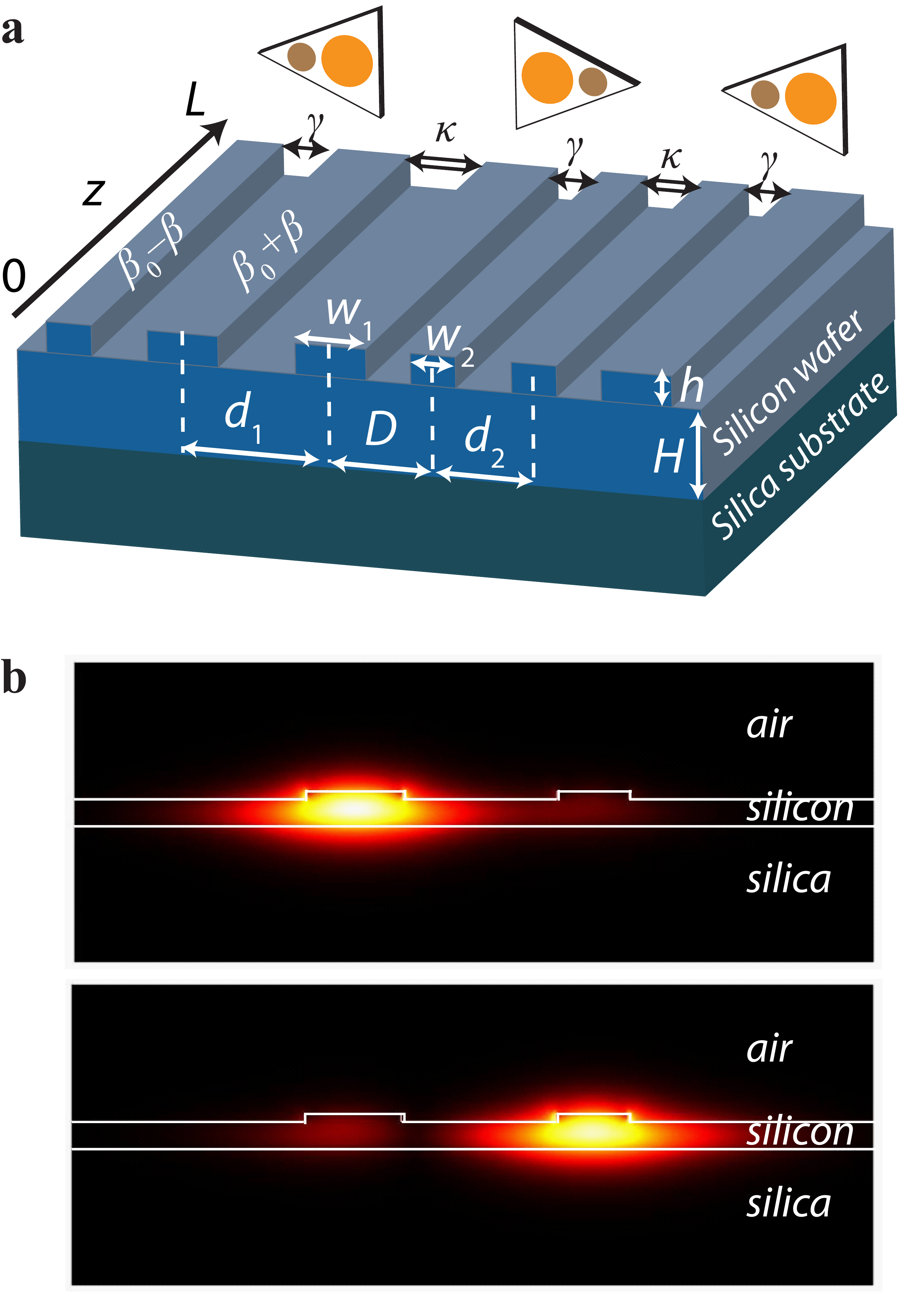}
	\caption{\label{fig6}
\textbf{Silicon photonics realization.} (a) The bow-tie chain can be realized in an integrated photonic structure made of silicon waveguides on top of a silica substrate.
Each dimer consists of two waveguides of different dimensions, giving rise to detuned propagation constants $\beta_0\pm\beta$, while the interdimer and intradimer couplings $\gamma$, $\kappa$ can be controlled via the spacings. These parameters can be inferred from the formation of supermodes in an isolated dimer, as shown in (b) for a system with $\beta_0 = 11.24\,\mu \mathrm{m}^{-1}$ and
$\beta = \gamma  = 0.06\,\mu \mathrm{m}^{-1}$ (for the geometric parameters see
Table \ref{tbl:table_1}).
}
\end{figure}

Topological photonics was incepted by considering the behaviour of photonic crystals under the influence of magneto-optical effects \cite{Haldane2008,Raghu2008}. Subsequently, a range of mechanisms to effectively break time-reversal symmetry were identified  \cite{Wang2008,Hafezi2011,Fang2012,Khanikaev2013,Lu2014},
and so were mechanisms to engineer a chiral symmetry \cite{Malkova2009,Atala2013,Poli2015} or a charge-conjugation symmetry
\cite{Poddubny2014}. The first experimental realizations utilized microwave structures  \cite{Wang2009}, followed by implementations at optical frequencies relying on platforms including resonator arrays  \cite{Hafezi2013}, waveguide lattices  \cite{Rechtsman2013}, and optical quantum walks  \cite{Kitagawa2012}.
In most cases  the design of these structures is based on tight-binding models, e.g., within a coupled-mode description of resonator or waveguide arrays \cite{Fan2003,Okamoto2006}.
The paradigmatic Su-Schrieffer Heeger model with staggered couplings has received particular attention on a large  variety of platforms, with works  directly probing the topological nature of the bands  \cite{Malkova2009,Atala2013,Ling2015}, also in the non-hermitian regime \cite{Rudner2009,Schomerus2013,Zhao2015,Zeuner2015}, and utilizing the protected defect states for phenomena such as mode selection \cite{Schomerus2013,Poli2015} and on-chip optical diodes \cite{El-Ganainy2015}. Staggered on-site energies have been implemented using optical lattices \cite{Atala2013} as well as
resonators or waveguides, e.g., to realize a photonic  analogue of a Peierls-spin chain \cite{Keil2013}.
Negative couplings can be obtained, e.g., by using auxiliary components that mediate the coupling  \cite{Keil2016}.

To implement the bow-tie chain, we first apply a $\mathbb{Z}_2$ gauge transformation to make all couplings positive. The desired gauge transformation takes the form of a basis change
\begin{equation}
\boldsymbol{\psi}_n\to    \left\{
               \begin{array}{ll}
                 \boldsymbol{\psi}_n, & \hbox{$n$ even,} \\
                  \sigma_z\boldsymbol{\psi}_n, & \hbox{$n$ odd,}
               \end{array}
             \right.
\end{equation}
after which the Bloch Hamiltonian \eqref{eq:bloch} becomes
\begin{equation}
\label{eq:bloch2}
H(k)=\left(\begin{array}{cccc} \beta &\kappa & 0 & \gamma e^{-ik} \\ \kappa & \beta & \gamma e^{ik} & 0
\\
0 & \gamma e^{-ik} & -\beta  & \kappa\\ \gamma e^{ik} & 0 &\kappa & -\beta
 \end{array}\right).
\end{equation}
This corresponds to alternating couplings $\kappa$, $\gamma$, $\kappa$, $\gamma$ that now are all positive, while the onsite energies still follow the sequence $\beta$, $\beta$, $-\beta$, $-\beta$; topological defects amount to irregularities in these sequences.

As in the general discussion, we focus on cases where the structure can be interpreted as a succession of asymmetric dimers, represented by the triangular elements in the pictorial description of Figs.~\ref{fig1}--\ref{fig5}.  In a given physical realization, the components that support the two fundamental dimer modes can have a variety of shapes; all that is required is the existence of two modes that are spectrally well isolated from the remaining modes. The effective tight-binding description employed here then follows from the application of standard coupled mode theory \cite{Okamoto2006}.

Here we consider the most straightforward implementation, the ridge waveguide geometry shown in Fig.~\ref{fig6}(a)
in which each dimer is formed by two distinct waveguides that operate in the single-mode regime. In this setup, the diagonal elements of the Hamiltonian represent propagation constants rather than energies. The onsite elements are  controlled via modifying the size or material composition of the guiding channels, while the coupling coefficients can be tuned by adjusting their spacings.
A benefit of this waveguide geometry is the possibility to study the wave dynamics along the structure (designated as the $z$ coordinate), which in coupled-mode theory is generated by $i\,d\boldsymbol{\psi}(z)/dz=H\boldsymbol{\psi}(z)$.

\begin{table}[t]
	\begin{tabular}{|c|r|}
		\hline
		parameter & value \\
		\hline
		$w_1$  &  $600\mathrm{nm}$   \\
		$w_2$  &  $450\mathrm{nm}$   \\
		$h$    &  $50\mathrm{nm}$    \\
		$H$    &  $180\mathrm{nm}$   \\
		$D$    &  $965\mathrm{nm}$   \\
		\hline
	\end{tabular}
	\caption{\label{tbl:table_1} Geometric parameters for a silicon dimer with $\beta_0 = 11.24\,\mu \mathrm{m}^{-1}$ and $\beta = \gamma  = 0.06\,\mu \mathrm{m}^{-1}$, giving rise to the supermodes shown in Fig.~\ref{fig6}(b).}
\end{table}

\begin{table}[t]
	\begin{tabular}{|c|c|c|c|}
		\hline
		center-to-center 					 & case (a) 						  & case (b) 						& case (c) \\
		separation distance 				 & $\kappa = 0.5\gamma$		  & $\kappa = \sqrt{2}\gamma$			& $\kappa = 2\gamma$ \\ \hline
		$d_1$								 & $1220\,\mathrm{nm}$		  & $870\,\mathrm{nm}$			 & $750\,\mathrm{nm}$		  \\
		$d_2$								 & $1270\,\mathrm{nm}$		  & $850\,\mathrm{nm}$			 & $700\,\mathrm{nm}$		  \\
		\hline
	\end{tabular}
	\caption{\label{tbl:table_2} Geometric parameters corresponding to three representative values of the interdimer coupling $\kappa$, for fixed dimer configurations with $\beta=\gamma= 0.06\,\mu \mathrm{m}^{-1}$ as given in Table \ref{tbl:table_1}.}
\end{table}

For a realistic modelling, we consider silicon ridge waveguides on a silica substrate, as depicted in Fig.~\ref{fig6}, which we characterize using a full-wave finite element method  \cite{Comsol}.
The waveguides are designed to support a single fundamental TE mode at a free-space wavelength of $\lambda_0 = 1.55\,\mu\mathrm{m}$.
The geometric parameters (height and width) of each waveguide as well as their separation distance are listed in Table \ref{tbl:table_1}.
These design parameters correspond to a detuning $\beta = 0.06\,\mu \mathrm{m}^{-1}$ that equals the intradimer coupling strength,
so that $\gamma = \beta$.
Figure \ref{fig6}(b) illustrates the supermode structure of an isolated dimer under the above conditions.  The field profiles of the supermodes are clearly asymmetric, with one supermode localized mainly in the left waveguide while the other resides in the right waveguide.

\begin{figure}
	\includegraphics[width=\columnwidth]{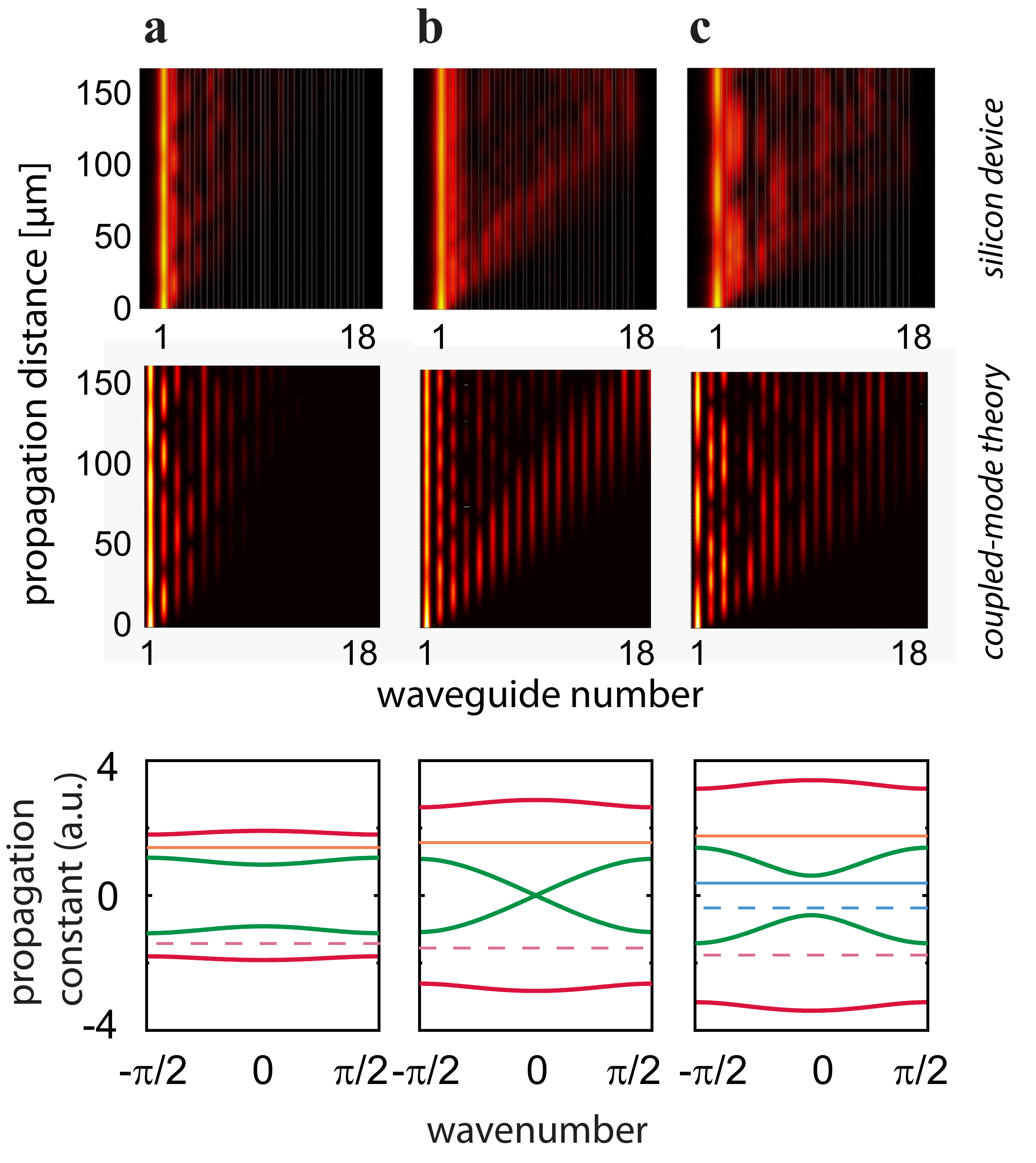}
	\caption{\label{fig7} \textbf{Probing edge states in wave propagation.}
The top panels show the predicted evolution of light intensity when an optical beam is launched into the leftmost waveguide of a silicon photonics structure of 18 waveguides, designed as shown in Fig.~\ref{fig6}(a). The dimers are configured to $\beta=\gamma=0.06\,\mu \mathrm{m}^{-1}$, while the interdimer coupling is set to (a) $\kappa=\gamma/2$, (b) $\kappa=\sqrt{2}\gamma$, and (c)  $\kappa=2\gamma$
(see geometric parameters in Tables \ref{tbl:table_1} and \ref{tbl:table_2} for the structure shown in Fig.~\ref{fig6}). The results of the  simulations coincide well with the predictions of couple-mode theory, shown in the middle panels. The  band structure underlying the couple-mode theory is shown in the bottom panels, with edge states at the left and right edge indicated by solid and dashed horizontal lines. In case (a), only a single edge state exists at the left edge [cf.~Fig.~\ref{fig4}(a)]; this state is clearly seen in the propagation. In case (b), the central gap closes, leading to a locally linear dispersion that gives rise to a characteristic diffraction pattern. In case (c), the gap is reopened, and the band inversion results in the formation of a second edge state  [cf.~Fig.~\ref{fig4}(c)]. Both edge states are clearly seen to interfere in the propagation intensity pattern.}
\end{figure}

For the extended system, we discuss three different array designs, giving rise to the representative band structures shown in the bottom panels of Fig.~\ref{fig7}. These scenarios are obtained by choosing the inter-dimer distances according to the values given in Table \ref{tbl:table_2}, which selects the interdimer coupling $\kappa$ while keeping $\beta=\gamma$ fixed.  In case (a) $\kappa=\gamma/2$, hence $\xi=\tilde\xi=1$, for which one topological edge state exists at the left side of the array (see  Fig.~\ref{fig4}(a); another such state  exists at the right edge). This edge state also exists in case (b) where $\kappa = \sqrt{2}\gamma$, at which still $\tilde\xi=1$ but the central bandgap is closed.  In case (c), $\kappa = 2\gamma$, so that now $\xi=-1$; then two localized modes having different eigenvalues and residing in different band gaps coexist at each edge (see Fig.~\ref{fig4}(c)).

As shown in the top panels of Fig.~\ref{fig7}, these modes can be probed by investigating the intensity evolution when an optical beam is launched into the leftmost waveguide. To mimic realistic experimental conditions, we consider an integrated silicon device made of 18 waveguides, each of which is $170\,\mu\mathrm{m}$ long.
In case (a), the intensity is guided by the edge state in the finite-frequency gap. In case (b), where the central gap is closed,  a secondary emission appears that resembles diffraction in uniform waveguide arrays \cite{Christodoulides2003}.
The subsequent appearance of a second localized mode in the central gap, case (c), can be inferred from the beating pattern of the optical intensity in the leftmost waveguide.
The results for the integrated silicon device agree well with those using coupled-mode theory, which are shown in the  middle panels.

The construction of the bow-tie chain, invoking a non-trivial square root,  is intimately linked to a tight-binding picture. The results in the present Section show that such considerations indeed transfer to realistic continuous systems, as long as they are suitably patterned to justify the coupled-mode description.

\section{Possible generalizations}
\label{sec:general}

\begin{figure*}[t]
	\includegraphics[width=1.5\columnwidth]{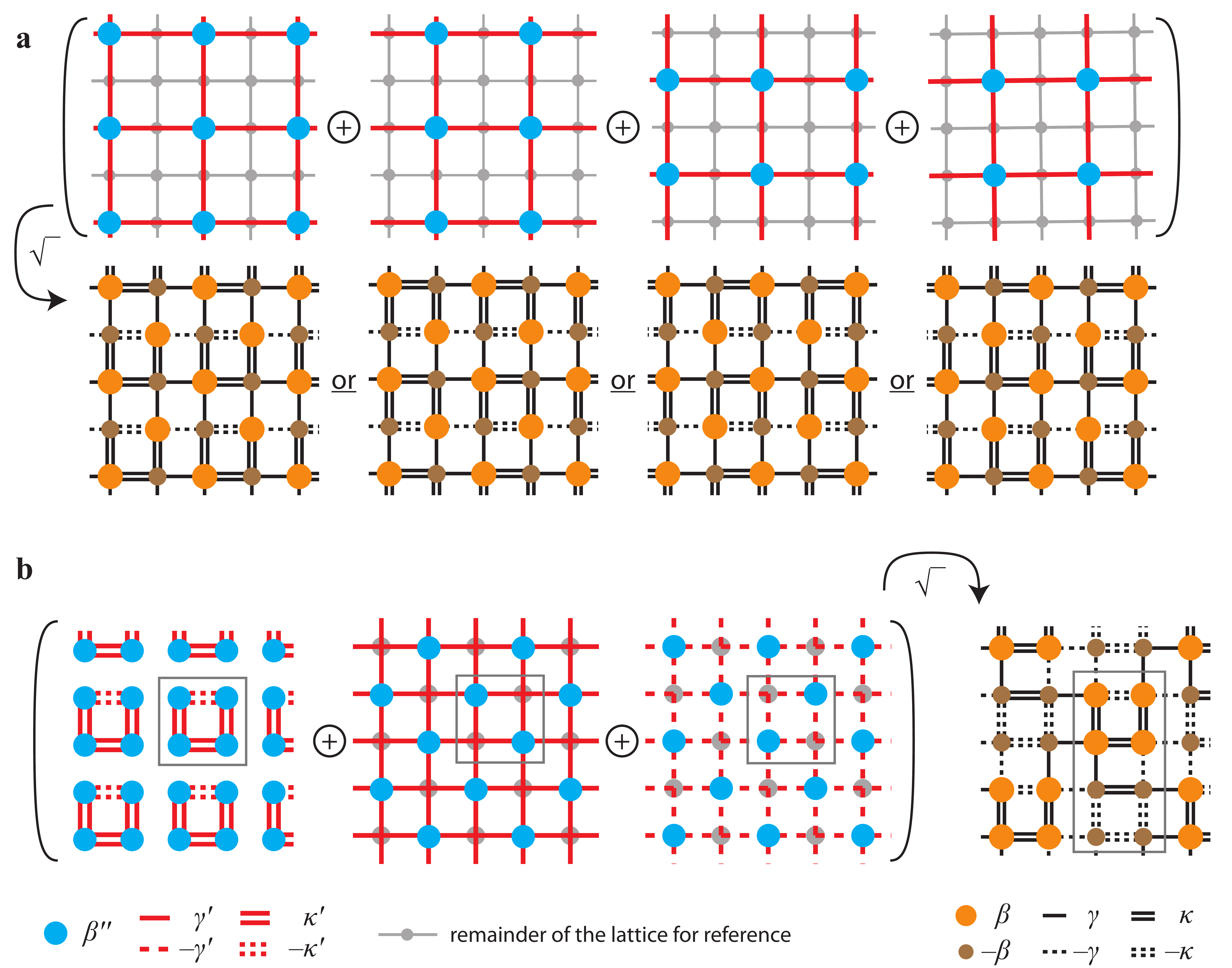}
	\caption{\label{fig8} \textbf{Extensions to two-dimensional square lattices}. (a) The $\pi$-flux square lattice, a systems which has gained considerable attention due to its topological and statistical features, can be interpreted as the square-root of four uncoupled square lattices. The square-root operation leaves freedom to choose between several dimerization patterns, whose interplay has been studied in the past to define topological defects with fractionalized charge \cite{Seradjeh2008}. Our construction reveals the additional freedom to choose an onsite corrugation pattern, which maybe exploited to define additional defects. While the unit cell retains four sites, we deem this square root non-trivial as the resulting system has a reduced rotation symmetry (this already applies without the onsite corrugation). The parameters of both models are  related by $\beta''=\beta^2+2\gamma^2+2\kappa^2$, $\gamma'=\gamma\kappa$.
(b) Starting from a square lattice system with $\pi$-fluxes in alternating cells, and additional next-nearest neighbour couplings (unit cell size 4), the square-root operation allows to obtain a system with 8 sites per unit cell. Again, there is freedom to choose between various dimerization and onsite corrugation patterns,  providing scope to form topological defects in the system. This system constitutes the most natural extension of the bow-tie chain into two dimensions. The couplings are defined in the same way as above, with in addition $\kappa'=2\beta\kappa$.}
\end{figure*}

As already mentioned in the discussion of Fig.~\ref{fig2}, the Rice-Mele model itself can be viewed as an example of a trivial square root, where the parent system consists of two uncoupled chains; both the square root and the parent system possess two sites per unit cell. Starting from the bow-tie chain, further generalization can be achieved by expanding the underlying algebra beyond the square-root operation. E.g., a spectrally shifted spectrum is obtained by starting from a parent tight-binding Hamiltonian written as $H'=H^2+\alpha H+\alpha'$, which we exploited implicitly for the photonic realization with a reference propagation constant $\beta_0$. In the construction of the bow-tie chain, we also restricted our attention to systems with couplings between adjacent unit cells. More remote couplings introduce non-monotonous bands, which is a prerequisite to generate multiple defect states in selected band gaps and realize the full scope of a topological quantum number $\nu\in\mathbb{Z}$. These features require the generalization
\begin{align} \label{eq:itergen}
&E^2\psi_n
=\left(H_n^2+\sum_{m\neq n}T_{nm}T_{mn}\right)\psi_n
\\&{}+\sum_{m\neq n}(H_nT_{nm}+T_{nm}H_m)\psi_m+\sum_{l\neq n;m\neq l,n}T_{nm}T_{ml}\psi_l,
 \nonumber
\end{align}
of the relations Eqs. \eqref{eq:iter1} and \eqref{eq:iter2} between the parent and the child system, obtained by iteration of a tight-binding equation with couplings $T_{ml}$  between cells $m$ and $l$. Additional bands can be created by increasing the period in the child system, so that the unit cell encompasses more components. Alternatively, one could shift the spectrum of the child to positive energies and take additional square roots, effectively generating polynomials of higher order.
The unifying key feature in this general one-dimensional setting is the observation that the square root operation allows to replace lattice symmetries by spectral symmetries, and that these can be interpreted as fractional lattice translations, as described in Sec. \ref{sec:prop}.

\begin{figure*}
	\includegraphics[width=1.6\columnwidth]{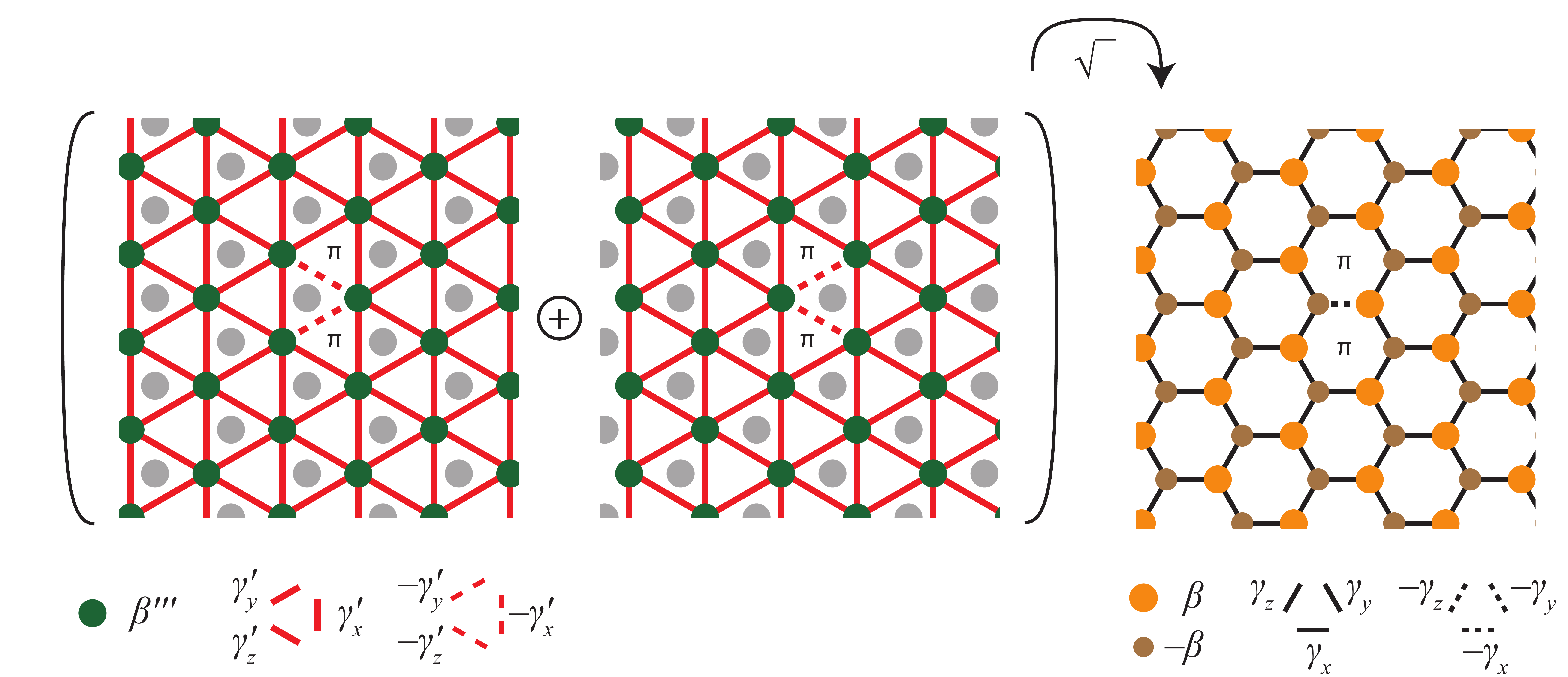}
	\caption{\label{fig9} \textbf{Anisotropic honeycomb lattice with $\pi$ fluxes.} The Kitaev honeycomb can be solved by a mapping to graphene-like single-particle sectors with anisotropic couplings and inverted bonds, where the latter generate an arrangement of $\pi$ fluxes \cite{Kitaev2006}. Each of these sectors can be viewed as a square root of two triangular lattices with a corresponding flux arrangement, as shown here for a pair of $\pi$ fluxes generated by a single inverted bond. The parameters in the parent system are $\beta'''=\beta^2+\gamma_x^2+\gamma_y^2+\gamma_z^2$ and $\gamma_i'=\gamma_x\gamma_y\gamma_z/\gamma_i$ with $i\in\{x,y,z\}$, where we also include a staggered on-site potential $\beta$ not present in the Kitaev honeycomb.}
\end{figure*}

To see how our considerations can be further extended beyond this one-dimensional setting, we now briefly describe  non-trivial examples in two dimensions, as illustrated in Figs.~\ref{fig8} and \ref{fig9}.
The first system [Fig.~\ref{fig8}(a)] highlights a typical feature of topologically non-trivial models in higher dimensions, namely, the role of  gauge configurations. When generated by a magnetic field, these configurations result in finite fluxes through plaquettes that cannot be gauged away. Starting from four uncoupled copies of square lattices without such fluxes (on-site energies $\beta''=\beta^2+2\gamma^2+2\kappa^2$ and couplings $\gamma'=\gamma\kappa$), a $\pi$-flux lattice can be generated via the square-root operation. Such lattices are characterized by one negative coupling around each plaquette, and
have been studied extensively due to their rich topological and statistical features (see, e.g., \cite{Mudry2003}). As we see in the figure, the square-root operation introduces the freedom to choose a dimerization pattern, as well as an on-site corrugation pattern. These features do not require to extend the unit cell (there are still four sites per unit cell), but break the rotational symmetry of the system; they therefore reduce the crystal symmetry. $U(1)$ and $Z_4$ defects in the dimerization pattern have been introduced in \cite{Seradjeh2008} to study charge fractionalization; working backwards as in Eqs.~\eqref{eq:iter1} and ~\eqref{eq:iter2}, such defects can now be seen to introduce localized potential variations and couplings between the four parent lattices. Further defects can be introduced through the onsite corrugation pattern, a feature which remains to be explored.

For the second example we generalize this system so that the square-root operation induces extra components. As shown in Fig.~\ref{fig8}(b), the parent system possesses four sites per unit cell and includes next-nearest-neighbour couplings, in analogy to the two-legged ladder. Choosing a suitable flux configuration and exploiting the freedom of dimerization and corrugation patterns, the non-trivial square root features a unit cell with 8 sites.
This system appears to be the most natural extension of the bow-tie chain to two dimensions.

As a third example, we consider honeycomb systems such as graphene \cite{CastroNeto2009}.
Anisotropic versions of these systems appear in the single-particle sector of the Kitaev honeycomb  \cite{Kitaev2006}, where $\pi$-fluxes are generated by couplings with an inverted sign (inverted bonds). As shown in Fig.~\ref{fig9}, the parent system is the sum of two triangular lattices, each with inverted bonds that generate a corresponding arrangement of $\pi$-fluxes. Symmetry breaking again occurs as a consequence of a spectral shift in the parent system, which under the square-root operation generates a sublattice-staggered potential $\pm\beta$. In absence of the fluxes, the resulting honeycomb system only displays a three-fold rotation symmetry about each plaquette centre, while the parent system displays a six-fold rotation symmetry.

These considerations suggest the following criterion to signify whether a square root is non-trivial---\emph{a trivial square root displays the same crystal symmetries as the parent system, while in a non-trivial square root some of these symmetries are broken down}. Supported by the examples studied here, and recalling also our
general considerations for one-dimensional systems in Section \ref{sec:prop}, we then argue that the reduction in symmetry gives scope for richer representations, which in the simplest case amounts to a larger unit cell, effectively equipping the system with more components.
The lost crystal symmetries may be replaced by new spectral constraints, as exemplified by  the chiral symmetry
\eqref{eq:chiraltb} in the bow-tie chain; additional features such as a chiral symmetry of the parent system yield further nontrivial constraints, as exemplified by Eq.~\eqref{eq:nonlinrelation}.

\section{Concluding remarks and outlook}
\label{sec:conc}

In this work we set out to explore whether interesting topological features can arise when one considers the concept of a square root of a Hamiltonian, as utilized by Dirac in the pursuit of relativistic quantum mechanics, and transfers it to the setting of periodic tight-binding lattices. We identified the simplest non-trivial one-dimensional example, the bow-tie chain, and found that it possesses a rich topological band structure, providing means to generate a versatile combination of edge and interface states. These features arise from spectral symmetries and topological indices that emerge under the square-root operation while some crystal symmetries are broken. The model can be implemented in suitably engineered photonic systems, not only in the integrated silicon photonic structures considered in Section \ref{sec:phot}, but also, for example, in plasmonic devices and atom-optical settings, which have been used to implement a wide range of tight-binding systems.  In the context of electronic transport, the model may, e.g., be viewed as a topologically nontrivial extension of the Rice-Mele model, corresponding to a conjugated polymer with a larger unit cell.

While we mainly employed this concept to identify a minimal model that can be easily implemented, it also brings a new perspective to a broad range of systems that are already under investigation. By working backwards as in Eqs.~\eqref{eq:iter1} and ~\eqref{eq:iter2}, it is indeed not difficult to relate a variety of well-known systems to simple parent configurations. Besides the $\pi$-flux square and honeycomb lattices described in Section \ref{sec:general}, this applies, e.g., to the Lieb lattice \cite{Lieb1989}
or the Haldane model \cite{Haldane1988}. We are certain that many rich examples of these correspondences remain to be discovered.

In summary, the utility of the constructions presented here is two-fold---by taking a non-trivial square root, it is possible to construct simple but topologically rich models from well-understood parent systems; by reading the relations backwards, they present a tool to gain additional insights into a  range of interesting models.
As an application, we constructed the bow-tie chain, a simple system with two topological invariants that can be implemented on a variety of platforms. Our considerations also open up a range of general questions --- in particular, concerning the precise scope for the non-trivial effects that can emerge, both from the point of view of representation theory as well as regarding their place in the well-established system of topological universality classes \cite{Ryu2010,Teo2010,Chiu2016,finalremark}.

\appendix

\begin{figure*}
	\includegraphics[width=1.6\columnwidth]{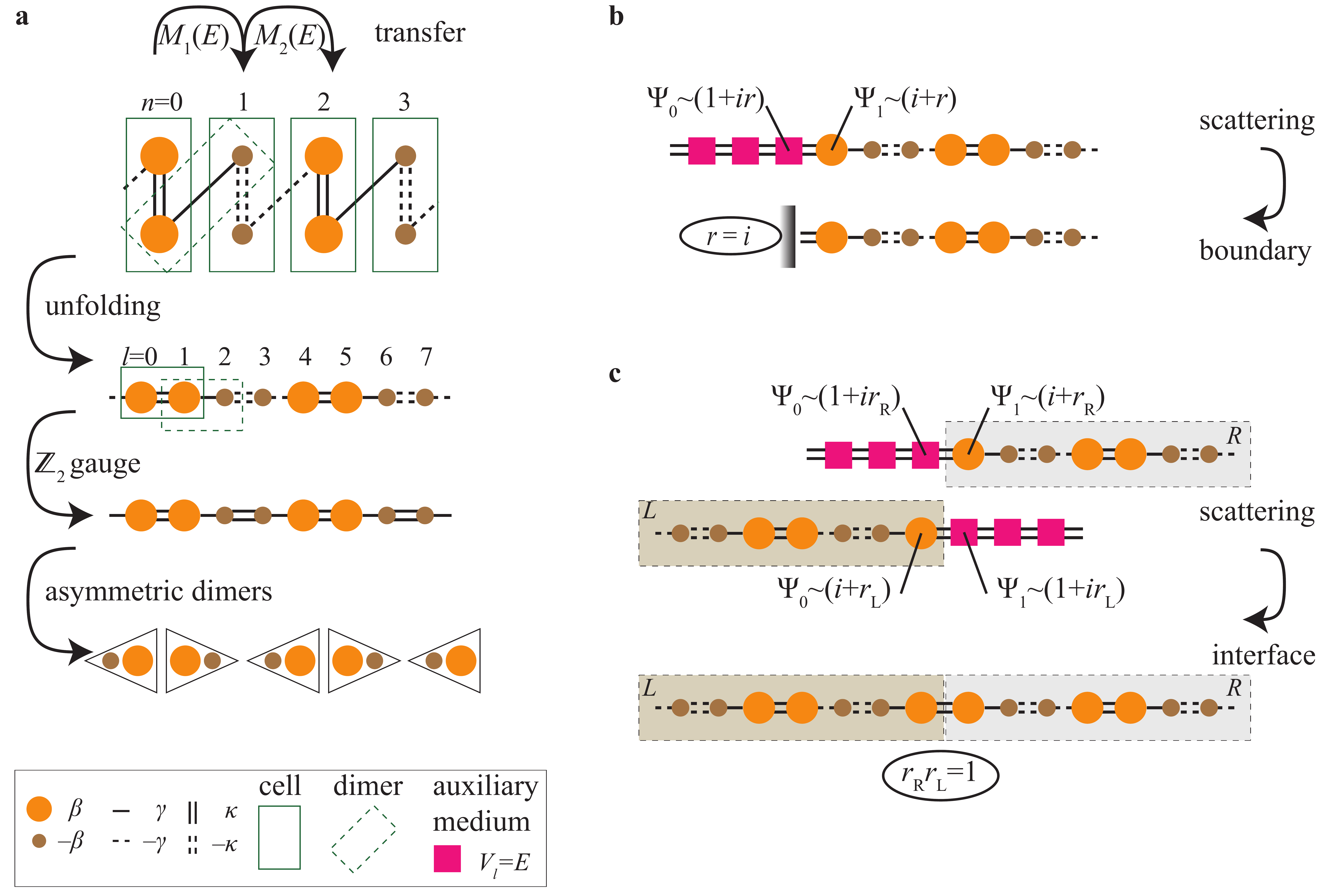}
	\caption{\label{figapp} \textbf{Transfer and scattering approach in the unfolded chain.} (a) Detailed definition of the unfolded chain and symmmetric dimers. The transfer matrices translate amplitudes between cells as defined in the top panel. These matrices can be inferred from the corresponding linear chain defined in the middle panel. After the $\mathbb{Z}_2$ transformation, the syste can be represented by two types of oppositely orientated asymmetric dimers (bottom panel). Note that the dimers (dashed boxes) span across two cells (solid boxes). (b) Definition of the reflection coefficient for scattering from a semi-infinite system. The external medium is perfectly matched, and its characteristics drop out when the system is terminated. This leads to the simple quantization condition for edge states $r(E)=i$. (c) The scattering approach also allows to address interfaces between two semi-infinite systems (denoted L and R), where the quantization condition takes the form $r_R(E)r_L(E)=1$. }
\end{figure*}

\section{Detailed consideration of the topological features}\label{app}
In this Appendix we provide the technical details of the topological characterization of the bow-tie chain, including the specific construction of the transfer and reflection matrices, the way these quantities enter the defect quantization condition and how the solutions of these conditions are linked to the topological indices. We also discuss the general topological classification of the system.

\subsection{Transfer matrix and scattering approach}\label{app1}
As shown in more detail in Fig.~\ref{figapp}(a), the bow-tie chain in Fig.~\ref{fig1} can be unfolded into a linear chain with nearest-neighbour couplings. As shown in Fig.~\ref{figapp}(b,c), this allows us to formulate a convenient scattering picture for the formation of defect states at boundaries and interfaces, which directly connects to the topological analysis of the band structure.

We start with the definition of the transfer matrix. Exploring the fact that we only have nearest-neighbour couplings, we can set $M=1$ in the tight-binding equations \eqref{eq:tb}. Denoting these amplitudes for clarity as $\Psi_l$, where in our previous notation $\Psi_{2n}=\psi_{n,1}$ and $\Psi_{2n+1}=\psi_{n,2}$ [see Fig.~\ref{figapp}(a)], they take the form
\begin{equation}
\label{eq:app1}
E\Psi_l=V_l\Psi_l+t_{l-1}\Psi_{l-1}+t_l\Psi_{l+1}.
\end{equation}
Here we assumed that the  coefficients $t_l$ are real, which can always be achieved by adopting a suitable gauge, i.e., fixing the phase of the amplitudes $\Psi_l$. We still retain the $\mathbb{Z}_2$ gauge freedom of choosing the signs of $\Psi_l$, which allows us to switch the sign of any coupling.
Given Eq.~\eqref{eq:app1}, we obtain
\begin{equation}
\label{eq:app2}
\Psi_{l+1}=\frac{E-V_l}{t_l}\Psi_l-\frac{t_{l-1}}{t_l}\Psi_{l-1},
\end{equation}
from which we read off the transfer matrix
\begin{equation}
\tilde M_{l}(E)=\left(\begin{array}{cc}0 & 1 \\ -t_{l-1}/t_l & (E-V_l)/t_l\end{array}\right).
\end{equation}
Given the amplitudes $\Psi_0$ and $\Psi_1$ on two neighbouring sites, the amplitudes throughout the system can then be inferred from
\begin{equation}
\left(\begin{array}{c}\Psi_{l} \\ \Psi_{l+1}\end{array}\right)=\tilde M_{l}(E)\times \tilde M_{l-1}(E)\times\ldots\times \tilde  M_1(E) \left(\begin{array}{c}\Psi_{0} \\ \Psi_{1}\end{array}\right).
\end{equation}
Reverting back to the original paired amplitudes within each cell, we thus have
\begin{equation}
\boldsymbol{\psi}_n=M_n(E)\times M_{n-1}(E)\times\ldots\times M_1(E) \boldsymbol{\psi}_0,
\end{equation}
where $M_n(E)=\tilde M_{2n}(E)\tilde M_{2n-1}(E)$.

In the linear chain, the couplings follow the repeating pattern  $(t_0,t_1,t_2,t_3)= (\kappa,\gamma,-\kappa,-\gamma)$,  while the onsite energies repeat according to $(V_0,V_1,V_2,V_3) = (\beta,\beta,-\beta,-\beta)$. Therefore,
\begin{subequations}
\begin{align}
\tilde M_1(E)&=\left(\begin{array}{cc}0 & 1 \\ -\kappa/\gamma & (E-\beta)/\gamma\end{array}\right),
\\
\tilde M_2(E)&=\left(\begin{array}{cc}0 & 1 \\ \gamma/\kappa & -(E+\beta)/\kappa\end{array}\right),
\\
\tilde M_3(E)&=\left(\begin{array}{cc}0 & 1 \\ -\kappa/\gamma & -(E+\beta)/\gamma\end{array}\right),
\\
\tilde M_4(E)&=\left(\begin{array}{cc}0 & 1 \\ \gamma/\kappa & (E-\beta)/\kappa\end{array}\right).
\end{align}%
\end{subequations}
The transfer by a period of the system (two cells) is then described by the product $M(E)=\tilde M_4(E)\tilde M_3(E)\tilde M_2(E)\tilde M_1(E)$.

As required by flux conservation, this matrix is symplectic, $M^\dagger(E)\sigma_y M(E)=\sigma_y$. The eigenvalues can be written as $\Lambda_\pm(E)=\exp(\pm ik(E))$ (hence $\Lambda_+(E)\Lambda_-(E)=1$), and determine the band structure of the system. For energies in the bands, $k(E)$ is real, hence $|\Lambda_+(E)|=|\Lambda_-(E)|=1$, while in the gaps we choose the sign of $k(E)=i\kappa(E)$ according to $\mathrm{Re}\,\kappa>0$, so that $|\Lambda_+(E)|<1$ describes an evanescent wave that decays to the right, while $|\Lambda_-(E)|>1$ describes an evanescent wave that decays to the left. The associated eigenvectors are denoted by $\boldsymbol{\phi}_\pm(E)$.

For the formation of an edge state in a semi-infinite system as shown in Fig.~\ref{fig4}, we require the boundary condition $\Psi_0=\psi_{0,1}=0$ [see Fig.~\ref{figapp}(b)]. This has to be compatible with an evanescent mode that decays to the right, hence
\begin{equation}
\phi_{+,1}(E)=0.
\label{eq:appbc1}
\end{equation}
Analogously, a boundary cutting across a dimer element enforces the boundary condition $\Psi_1=\psi_{0,2}=0$, hence
\begin{equation}
\phi_{+,2}(E)=0.
\label{eq:appbc2}
\end{equation}
For the interfaces shown in Figs.~\ref{fig1}(a) and \ref{fig5}(a,b), we require that decaying modes in the left and right medium match up as shown in Fig.~\ref{figapp}(c), which leads to the condition
\begin{equation}
\boldsymbol{\phi}_-^L(E)=\boldsymbol{\phi}_+^R(E).
\label{eq:appbc3}
\end{equation}

To set up the scattering formulation of these conditions, we consider a semi-infinite chain attached to an exterior medium [see again Figs.~\ref{figapp}(b,c)]. The  details of this medium will drop out once we replace it by a boundary or an interface to another system. We  therefore assume an ideally matched featureless medium, which is obtained by continuing the system as a monoatomic chain with onsite potential set to the energy of the system. We denote the amplitudes in the medium as $\Psi_l$ with $l\leq 0$, while for the system $l\geq 1$. In the medium, the solution is a Bloch wave of the form $\Psi_l=A(e^{ik_0(l-1/2)} +re^{-ik_0(l-1/2)})$, where $k_0$ =$\pi/2$ due to our choice of the onsite potential, while the offset by $1/2$ refers the reflected wave to the effective location of the interface. This wave has to match with the decaying wave in the system, which is achieved if
\begin{subequations}
\begin{align}
\phi_{+,1}(E)&=A(e^{ik_0(-1/2)} +r(E)e^{-ik_0(-1/2)})
\nonumber\\
&=A e^{-i\pi/4}(1+ir(E)),\\
\phi_{+,2}(E)&=A(e^{ik_0(1/2)} +r(E)e^{-ik_0(1/2)})\nonumber\\
&=A e^{-i\pi/4}(i+r(E)).
\end{align}%
\end{subequations}
From this we obtain the reflection coefficient $r(E)$ as given in Eq.~\eqref{eq:refl}.

\subsection{Quantization conditions and topological invariants}\label{app2}

In terms of this reflection coefficient, the boundary condition \eqref{eq:appbc1} yields the quantization condition $r(E)=i$ while the boundary condition \eqref{eq:appbc2} yields the quantization condition $r(E)=-i$.
To describe an interface, we analogously define
\begin{subequations}
\begin{align}
r_R(E)&=\frac{\phi_{+,1}^R(E)+i\phi_{+,2}^R(E)}{\phi_{+,2}^R(E)+i\phi_{+,1}^R(E)},
\\
r_L(E)&=\frac{\phi_{-,2}^L(E)+i\phi_{-,1}^L(E)}{\phi_{-,1}^L(E)+i\phi_{-,2}^L(E)}.
\end{align}%
\end{subequations}
The boundary condition \eqref{eq:appbc3} then requires $r_R(E)r_L(E)=1$, which can be interpreted as the condition for constructive interference in a round trip through the system.

Within the gaps, the reflection coefficient $|r(E)|=1$ and the winding of its phase is constrained by the symmetry-constrained values $r=\pm 1$ at the different band edges. It follows that the winding is a topological invariant and can only change when band gaps are closed. The number of solutions for the various boundary conditions can therefore be inferred from the topological features of the bandstructure.

To characterize these features, we first consider the Zak phase $Z$, defined in Eq.~\eqref{eq:zak}.
As a first step, let us parameterize
\begin{equation}
\boldsymbol{\varphi}(k)=N(k)\left(\begin{array}{c} 1\\ \alpha(k) \\ A(k)\\  A(k)\tilde \alpha(k)\end{array}\right)
\end{equation}
with normalization constant $N(k)$  and parameters $\alpha(k)$, $\tilde\alpha(k)$, and $A(k)$.
The symmetries \eqref{eq:symm} imply $\boldsymbol{\varphi}(k)\propto \mathcal{R}\boldsymbol{\varphi}^*(k)$,
from which we obtain the constraints $\tilde \alpha(k)=-A^*(k)\alpha(k)/A(k)$ as well as $|\alpha(k)|^2=1$. Hence
\begin{equation}
\label{eq:param}
\boldsymbol{\varphi}(k)=N(k)\left(\begin{array}{c} 1\\ \alpha(k) \\ A(k)\\  -A^*(k) \alpha(k)\end{array}\right),\,\, N(k)=\frac{1}{\sqrt{2(1+|A(k)|^2)}}.
\end{equation}

To further exploit the symmetries \eqref{eq:symm}, we split the integral into the right-propagating branch $\boldsymbol{\varphi}_+(E)$ and the left-propagating branch $\boldsymbol{\varphi}_-(E)=\boldsymbol{\varphi}_+^*(E)$, parameterised by energy. The Zak phase then takes the form
\begin{align}
z&=
i\,{\rm Im}\,\int_{E_{min}}^{E_{max}}dE\,
[\boldsymbol{\varphi}_+^\dagger(E)\frac{d}{dE}\boldsymbol{\varphi}_+(E)
-\boldsymbol{\varphi}_-^\dagger(E)\frac{d}{dE}\boldsymbol{\varphi}_-(E)]
\nonumber
\\
&=-2\,{\rm Im}\,\int_{E_{min}}^{E_{max}}dE\,
\boldsymbol{\varphi}_+^\dagger(E)\frac{d}{dE}\boldsymbol{\varphi}_+(E)
\nonumber
\\
&=-{\rm Im}\,\int_{E_{min}}^{E_{max}}dE\, \alpha_+^*(E)\frac{d}{dE}\alpha_+(E),
\end{align}
where the band edges $E_{min}$ and $E_{max}$ are taken from Eq.~\eqref{eq:edges}, while in the last step we made use of the parametrization \eqref{eq:param}. It follows that the Zak phase is determined by the winding of $\alpha_+(E)$.

The first two components of the Bloch wave $\boldsymbol{\varphi}_+(E)$ coincide with the eigenvector $\boldsymbol{\phi}_+(E)$ of the transfer matrix. Therefore, the reflection coefficient is
\begin{equation}
r(E)=\frac{1+i\alpha_+(E)}{\alpha_+(E)+i},
\mbox{ hence }
\alpha_+(E)=\frac{r(E)+i}{1+ir(E)}.
\end{equation}
The constraint $|\alpha_+(E)|^2=1$ implies that within a band $r(E)$ is real, while $|r(E)|\leq 1$ implies that $\mathrm{Im}\,\alpha_+(E)\geq 0$. It follows that multiple windings of this phase factor are forbidden, allowing us to fix the branch $0\leq\mathrm{arg}\,\alpha_+(E)\leq \pi$. With these constraints, it suffices to know $\alpha_+(E)$ at the band edges,
\begin{equation}
z=\mathrm{arg}\,\alpha_+(E_{min})-\mathrm{arg}\,\alpha_+(E_{max}).
\end{equation}
As at the edges the reflection coefficient can only take the values $r(E_{min,max})=\pm1$, we have $\mathrm{arg}\,\alpha_+(E_{min,max})=[1+r(E_{min,max})]\pi/2$, from which we recover Eq.~\eqref{eq:zak2}.
More formally, this result can be substantiated by the transformation
\begin{align}
z&=-{\rm Im}\,\int_{E_{min}}^{E_{max}}dE\, \alpha_+^*(E)\frac{d}{dE}\alpha_+(E)
\nonumber
\\
&=-2\int_{E_{min}}^{E_{max}}dE\, \frac{1}{1+r^2(E)}\frac{d}{dE}r(E)
\nonumber
\\
&=2\arctan r(E_{min})-2\arctan r(E_{max}).
\end{align}
Given that $\arctan(\pm1)=\pm\pi/4$ (with other branches not accessible due to the constraint that $r(E)$ is real and obeys $| r(E)|\leq 1$), we again recover Eq.~\eqref{eq:zak2}.

The explicit value of the reflection coefficient at a given band edge $E_0=E_{min},E_{max}$, specified again by Eqs.~\eqref{eq:edges}, is most quickly obtained by testing whether the transfer matrix $M(E_0)$ has an eigenvector $(1,1)^T$ or $(1,-1)^T$. This amounts to the conditions
\begin{subequations}
\begin{align}
(1,-1)M(E_0)\left(\begin{array}{c} 1\\ 1\end{array}\right)=0 &\quad \mbox{for }r(E_0)=1,
\\
(1,1)M(E_0)\left(\begin{array}{c} 1\\ -1\end{array}\right)=0 &\quad \mbox{for }r(E_0)=-1,
\end{align}
\label{eq:edgeref}%
\end{subequations}
and leads to the Zak phases as summarized in Eq.~\eqref{eq:zakresult}.

These considerations also allow us to establish the winding of the reflection coefficient in the gaps.
For each gap, we find from the conditions \eqref{eq:edgeref} that the reflection coefficients at the upper and lower band edge are always opposite.
We also find that $r(E)=\pm1$ only occurs at these edges, and at no other energies in the system.
From the conditions
\begin{subequations}
\begin{align}
(1,0)M(E')\left(\begin{array}{c} 0\\ 1\end{array}\right)=0 &\quad \mbox{for }r(E')=i,
\\
(0,1)M(E')\left(\begin{array}{c} 1\\ 0\end{array}\right)=0 &\quad \mbox{for }r(E')=-i,
\end{align}
\label{eq:ri}%
\end{subequations}
we analogously always find exactly one value $r(E')=\pm i$ of the reflection coefficient within each of the gaps. These values fix the winding of $r(E)$ in the gap, according to the Witten index as summarized in Eq.~\eqref{eq:wittenres}. In combination with the values at the edges, we further verify that in each gap, the winding of the reflection coefficient is always by $\pi$ and occurs in the clockwise sense. This agrees with the scenarios illustrated in Fig.~\ref{fig3}.

\subsection{Universality class}\label{app3}

As mentioned in the main text, given the symmetries of the bow-tie chain it is reasonable to place this system into the BDI symmetry class of chiral systems with a conventional time reversal symmetry. For the finite energy gaps, the analogous conclusion follows from the consideration of the parent systems, implying also that the topological features in the two finite-energy gaps are related.  As we have shown in detail in the previous section of this Appendix, the fact that we can still generate different states in these finite-energy gaps can be understood from the specific quantization  conditions at interfaces and boundaries. In practice, we find at most one defect state in each gap, but this can be attributed to the restriction to couplings between neighbouring unit cells, in  analogy to the situation in the SSH model (the general considerations in Sec. \ref{sec:prop} do not rely on these assumption; for a starting point for richer one-dimensional models see Eq. \eqref{eq:itergen}). What requires a more careful consideration, however, is the fact that the chiral symmetry in the bow-tie chain constitutes a fractional lattice translation, and therefore is nonsymmorphic. In principle, this can modify the topological nature of the central gap. We therefore supplement the topological classification from the perspective of systems with an additional order-two lattice symmetry $S$, characterised by the feature $[S^2,H(k)]=0$, for which a complete classification has been developed in Ref. \cite{Shiozaki2014}. We make direct use of the results in the cited work, in the hope that they are of interest for the specialist reader (equation and table numbers in the following refer to the cited work).

Ref. \cite{Shiozaki2014} distinguishes between order-two lattice symmetries that are unitary or antiunitary and commute or anticommute with the Bloch Hamiltonian or other symmetries, in particular time-reversal symmetry. We start in symmetry class AI ($s=0$) for systems with a conventional time-reversal symmetry. Our chiral symmetry $X$ fulfills the criteria for a unitary symmetry that anticommutes with the Hamiltonian, squares to $+1$ and commutes with the time-reversal symmetry; such operations are denoted as $\bar U^+_+$.
According to Table IV we are therefore concerned with the case
$t=3$. The chiral symmetry $X$ conserves the momentum, so that $d_\parallel=0$, $d_\perp=d=1$ [see Eq. (3.16) for the decomposition of directions], and the defect dimensions
are $D=D_\parallel=0$. According to equations (3.27) and (3.31), $K^U_{\mathbb{R}}(s=0,t=3,d=1,d_\parallel=0,D=0,D_\parallel=0)=K^U_\mathbb{R}(-1 \mod 8,3,0,0,0,0)=\pi_0(\mathcal{R}_0)$.
Table I then confirms that the topological invariant takes values in $\mathbb{Z}$, which indeed coincides with class BDI in $d=1$ dimensions.

Finally, we remark that there
is a striking duality between these considerations, which are intimately linked to the commutation relation
$[X^2,H]=0$ (interpreting $X$ as a chiral symmetry linked to an order-two spatial symmetry), and our considerations, which are intimately linked to the relation $[X, H^2]=0$ (starting from a spatial symmetry of a parent Hamiltonian $H^2$, which generates a chiral symmetry for the underlying child Hamiltonian $H$).
Indeed, it is simple to show that, for general systems,
anti-commutative and commutative symmetries represent
commutative symmetries in even powers of either the
Hamiltonian or the symmetry operation itself. The physical
ramifications of this are left as an open question.


\begin{thebibliography}{99}
\bibitem{remark_flat_bands} Flat bands arise when the chiral operators has unequal numbers of eigenvalues $\pm1$, hence $\mathrm{tr}\mathcal{X}\neq 0$.
\bibitem{Comsol} The  calculations of wave propagation in the silicon ridge waveguides were carried in COMSOL using the full-wave finite element method.
\end{thebibliography}
\end{document}